\documentclass[a4paper,11pt]{article}
\pdfoutput=1
\usepackage{jheppub}
\usepackage{bbm,bm,graphicx,mathtools,color,hyperref,slashed}
\usepackage{amssymb,amsmath}
\usepackage[T1]{fontenc}

\usepackage[all]{xy}

\graphicspath{{./Figures/}}

\newcommand{\diff}{\mathrm{d}}
\newcommand{\p}{\partial}

\newcommand{\Diff}{{\mathcal{D}}}

\newcommand{\tr}{\mathrm{tr}}
\newcommand{\im}{\mathrm{i}}

\newcommand{\calO}{\mathcal{O}}

\newcommand{\calN}{\mathcal{N}}

\newcommand{\calZ}{\mathcal{Z}}

\newcommand{\rme}{\mathrm{e}}

\usepackage{comment}

\preprint{YITP-22-05}

\title{Exploring the $\theta$-vacuum structure in the functional renormalization group approach}

\author[1]{Kenji Fukushima,}
\emailAdd{fuku@nt.phys.s.u-tokyo.ac.jp}
\author[1]{Takuya Shimazaki,}
\emailAdd{shimazaki@nt.phys.s.u-tokyo.ac.jp}
\affiliation[1]{Department of Physics, The University of Tokyo, 7-3-1 Hongo, Bunkyo-ku, Tokyo 113-0033, Japan}
\author[2]{and Yuya Tanizaki\,}
\affiliation[2]{Yukawa Institute for Theoretical Physics, Kyoto University, Kyoto 606-8502, Japan}
\emailAdd{yuya.tanizaki@yukawa.kyoto-u.ac.jp}

\abstract{
We investigate the $\theta$-vacuum structure and the 't~Hooft anomaly at $\theta=\pi$ in a simple quantum mechanical system on $S^1$ to scrutinize the applicability of the functional renormalization group (fRG) approach.
Even though the fRG is an exact formulation, a naive application of the fRG equation would miss contributions from the $\theta$ term due to the differential nature of the formulation.
We first review this quantum mechanical system on $S^1$ that is solvable with both the path integral and the canonical quantization.
We discuss how to construct the quantum effective action including the $\theta$ dependence.  Such an explicit calculation poses a subtle question of whether a Legendre transform is well defined or not for general systems with the sign problem.
We then consider a deformed theory to relax the integral winding by introducing a wine-bottle potential with the finite depth $\propto g$, so that the original $S^1$ theory is recovered in the $g\to\infty$ limit.
We numerically solve the energy spectrum in the deformed theory as a function of $g$ and $\theta$ in the canonical quantization.
We test the efficacy of the simplest local potential approximation (LPA) in the fRG approach and find that the correct behavior of the ground state energy is well reproduced for small $\theta$.
When the energy level crossing is approached, the LPA flow breaks down and fails in describing the ground state degeneracy expected from the 't~Hooft anomaly.
We finally turn back to the original theory and discuss an alternative formulation using the Villain lattice action.
The analysis with the Villain lattice at $\theta=\pi$ indicates that the nonlocality of the effective action is crucial to capture the level crossing behavior of the ground states.
}

\begin{document}
\maketitle

\section{Introduction}

Renormalization group (RG) \`a la Wilson is a versatile method to interpolate the ultraviolet (UV) and infrared (IR) scales of quantum field theories (QFTs)~\cite{Wilson:1973jj,Wilson:1974mb,Polchinski:1983gv,Pelissetto:2000ek}, which gives theoretical understanding of the universality in critical phenomena.
The functional renormalization group (fRG) is a nonperturbative formulation of the RG flow and its applicability is not limited to the critical point~\cite{Berges:2000ew,Bagnuls:2000ae,Polonyi:2001se,Gies:2006wv,Dupuis:2020fhh}.
In fact the fRG could be regarded as an exact formulation of QFTs based on the functional differential equation~\cite{Delamotte:2003dw,Schaefer:2006sr,Metzner:2011cw,Canet:2011ez,Leonhardt:2019fua}.
This viewpoint contrasts with the conventional definition of QFTs formulated with functional integration.
There are several equivalent formulations of the fRG; namely, the Polchinski equation~\cite{Polchinski:1983gv}, the Wegner-Houghton equation~\cite{PhysRevA.8.401}, and the Wetterich equation~\cite{Wetterich:1989xg,Wetterich:1992yh}.
The key ingredient for the fRG equation is the effective average action $\Gamma_k$ that flows with the RG scale $k$, where $\Gamma_k$ is a coarse-grained action over length scale $k^{-1}$.
Therefore, the fRG equation, particularly the Wetterich equation that we employ in this work, makes interpolation between the known UV action, $\Gamma_\Lambda=S$, and the IR effective action, $\Gamma_0=\Gamma$.
We should emphasize that the fRG equation by itself is exact without any approximation, and one can thus define the theory nonperturbatively with the fRG equation and the initial condition; $\Gamma_\Lambda=S$.
There are a countless number of successful fRG applications such as the
$O(N)$ models~\cite{Bohr:2000gp,Defenu:2017dec,Connelly:2020gwa,DePolsi:2020pjk,DePolsi:2021cmi,Rose:2021zdk},
the Quark-Meson model~\cite{Schaefer:1999em,Sasaki:2006ww,Sasaki:2006ws,Schaefer:2007pw,Schaefer:2009ui,Herbst:2010rf,Herbst:2013ail,Kamikado:2013pya,Andersen:2013swa,CamaraPereira:2020xla},
the Nambu-Jona-Lasinio mode~\cite{Braun:2019aow},
quantum chromodynamics (QCD)~\cite{Braun:2009gm,Christiansen:2014ypa,Fu:2019hdw,Fu:2021oaw},
the quantum gravity~\cite{Litim:2003vp,Niedermaier:2006ns,Reuter:2007rv,Eichhorn:2018yfc},
etc.
It is simply impossible to mention all the progresses here, and the fRG prospects are further expanding.
In practice, we usually have to adopt some truncation to solve $\Gamma_k$ approximately.
There are several truncation schemes on $\Gamma_k$ suited for various purposes~\cite{Blaizot:2005xy,Blaizot:2005wd,Benitez:2011xx}.
One commonly used truncation is based on the derivative expansion~\cite{Morris:1999ba,Litim:2001dt,Balog:2019rrg}.

Even though the fRG is a promising theoretical tool to explore QFTs and their phase diagrams, it seems that current success is limited to the cases with local order parameters.
As mentioned above, the fRG is formulated by the functional differential equation, and it is a nontrivial question whether the formulation can correctly capture global properties of QFTs.
Global properties, especially the topology of the field space, play essential roles to understand quantum phases and quantum phase transitions that are beyond the Landau-Ginzburg paradigm~\cite{Haldane:1983ru,Haldane:1988zz,PhysRevB.70.144407}.

To see the subtlety of fRG in a more concrete shape, let us consider QFTs with the topological $\theta$ term.
The $\theta$ term appears if the field space is classified by the topological charge $w\in\mathbb{Z}$, and it gives the phase factor $\exp(\im \theta w)$ when summing over the topological sectors in the functional integral.
As the winding number $w$ is integrally quantized, such a term does not affect the equation of motion.
Nevertheless, the ground state properties can depend on $\theta$, and moreover there can exist phase transitions with increasing $\theta$~\cite{Witten:1980sp, DiVecchia:1980yfw, Witten:1998uka} (see Ref.~\cite{Vicari:2008jw} for a review).
However, a finite $\theta$ causes the sign problem to the first-principles Monte Carlo simulation; see Refs.~\cite{Bonati:2013tt,Bonati:2019kmf,Gattringer:2018dlw,Sulejmanpasic:2019ytl,Gattringer:2019yof,Hirasawa:2020bnl} for recent attempts.
Therefore, it would be desirable to establish an alternative to the Monte Carlo simulation, and one of the possibilities to evade the sign problem would be the fRG approach.
Because $w$ is topological, however, the fRG equation seems to be insensitive to the topological $\theta$ term at all.
How can we explore the $\theta$-vacuum properties in the fRG method?
This is the central question we would like to address in this work.

Let us stress that the treatment of the topological $\theta$ term in the fRG is not a mere technical problem.
As we explain later, recent years have seen intriguing developments in the application of 't~Hooft anomaly matching to constrain the IR behaviors of QFTs~\cite{tHooft:1979rat,Frishman:1980dq,Coleman:1982yg,Kapustin:2014lwa,Kapustin:2014zva}
(see Refs.~\cite{Gaiotto:2017yup, Gaiotto:2017tne,Tanizaki:2017bam, Komargodski:2017dmc,Komargodski:2017smk, Tanizaki:2017qhf, Yamazaki:2017ulc, Tanizaki:2018xto, Misumi:2019dwq, Yonekura:2019vyz, Chen:2020syd, Honda:2020txe,Tanizaki:2022ngt} for recent studies on the $\theta$ term in gauge theories).
The 't~Hooft anomaly is an obstruction in gauging a global symmetry
and it strongly constrains the possible phases of matter in strongly coupled theories because of its RG invariance.
When the UV theory has an 't~Hooft anomaly, the IR effective theory must reproduce the same anomaly, and thus the IR theory is prohibited to be trivially gapped.
In the seminal work~\cite{Gaiotto:2017yup}, the $SU(N)$ Yang-Mills theory at $\theta=\pi$ is shown to have a nonperturbative 't~Hooft anomaly, which is a consequence of the subtle topological property of the $SU(N)$ gauge fields.
It is thus important to investigate whether the fRG is capable of describing topological features beyond the perturbative regime.
We note that there are already some previous works in which topological properties have been studied successfully with the fRG, e.g.\ in the sine-Gordon model~\cite{Nandori:2009ad,Nagy:2009pj,Pangon:2010uf,Bacso:2015ixa,Daviet:2018lfy}. However, these models do not have the $\theta$ term, so it would be worthwhile to study the topological aspects using the fRG in more detail.

In this paper, we choose the simplest model for studying how to cope with the topological $\theta$ term in the fRG approach.
To this end, we shall employ a quantum mechanical system on a circle $S^1$ within the framework of the fRG equation.
Despite its simplicity, the $S^1$ quantum mechanics without the potential term is nontrivial enough to accommodate the 't~Hooft anomaly and the topological $\theta$ term as closely discussed in Ref.~\cite{Gaiotto:2017yup} (see also Refs.~\cite{Kikuchi:2017pcp,Sueishi:2021xti}).
The ground state in the $S^1$ quantum mechanics is doubly degenerate at $\theta=\pi$, and it is a consequence of 't~Hooft anomaly matching as we shall discuss.
Usually, quantum tunneling resolves the ground-state degeneracy, or level crossing, in quantum mechanics, but the 't~Hooft anomaly tells that this is not the case for the $S^1$ quantum mechanics as the degeneracy is between the states with different $U(1)$ charges.
This model gives a prototype for the quantum phase transition between different symmetry-protected topological (SPT) states in higher-dimensional QFTs.
Let us make another remark before explaining our fRG results.
The $S^1$ quantum mechanics is a solvable and well-understood problem, and we review its solution both from the canonical quantization and the path integral in Sec.~\ref{sec:review_qm}.
Besides, we will compute the quantum effective action for this $S^1$ quantum mechanical system, which has not been done before, and this result is useful as a benchmark for the fRG calculation.
The by-product from this explicit computation is a clear recognition of the remnant of the sign problem that obscures the existence of the quantum effective action.

As we discussed, as long as $\theta w$ is topological, there is no way to introduce the $\theta$ dependence in the fRG equation.
There are essentially two strategies to overcome this problem.
One is to deform $\theta$ so that $\theta$ can have spacetime dependence as well as the scale $k$ dependence.
Along these lines, the flow of $\theta$ was considered, and it was found that the flow has UV and IR discontinuities~\cite{Reuter:1996be}.
Due to those complications, we would pursue another strategy; that is,
we embed the target space $S^1$ to $\mathbb{R}^2$ with a wine-bottle potential with the finite depth $\propto g$.
For $g<\infty$, $w$ no longer takes an exact integer, so the fRG equation can have $\theta$ dependence as desired.
We can recover the original theory by studying the limit of $g\to\infty$.
Importantly, we can still solve this modified theory at finite $g$ numerically in the canonical quantization.
We find the energy spectrum similar to the original one labeled by integral numbers.
We numerically confirm that the local potential approximation (LPA) works well to reproduce the correct $\theta$ dependence of the ground state energy up to a $\theta$ value where the energy level crossing occurs.
Thus, our idea to relax the topology with an enlarged target space has turned out to be effective, but the ground state degeneracy hinders our method.

We finally consider a more rigorous treatment of the $S^1$ quantum mechanics on the lattice to see why the LPA fails at the degenerate point.
For this we utilize the Villain lattice formulation~\cite{Villain:1974ir,Gattringer:2018dlw,Sulejmanpasic:2019ytl,Gattringer:2019yof}.
Interestingly enough, the energy spectrum in the Villain lattice formulation is identical to the one in the original continuum theory.
Careful examination of the quantum effective potential at $\theta=\pi$ clarifies that the level crossing of the ground states causes severe nonlocality in the effective action, which poses a serious question about the practical applicability of LPA-type approximations of the fRG beyond the level crossing point.
It is an important future study to think of a controllable nonlocal ansatz of the effective action to tackle this problem.

\section{Quantum mechanics on \texorpdfstring{$S^1$}{S1}}
\label{sec:review_qm}

In Secs.~\ref{sec:spectrum_S1} and \ref{sec:pathintegral_S1}, we give a brief review on quantum mechanics of a free particle living on a circle $S^1$.
This model enjoys the topological $\theta$ parameter, which affects the energy spectrum.
In Sec.~\ref{sec:spectrum_S1} we quantize the model in a canonical way to discuss the level crossing at $\theta=\pi$, where a first-order phase transition occurs.
We identify the origin of degenerated eigenenergies from the symmetry algebra.
Since the fRG is based on the functional integral, in Sec.~\ref{sec:pathintegral_S1}, we confirm that the same energy spectrum is derived from the Euclidean path-integral formulation.
In Sec.~\ref{sec:effective_action_S1}, as a guiding reference for the functional approaches, we shall explicitly construct the quantum effective action and discuss how to read off nontrivial properties of the energy spectrum from the effective action.
We will demonstrate that a subtle change of the quantum effective action is crucial for the underlying mechanism of the phase transition in this quantum mechanical system.

\subsection{Energy spectrum in the canonical quantization}
\label{sec:spectrum_S1}

\begin{figure}
    \centering
    \includegraphics[width=0.7\textwidth]{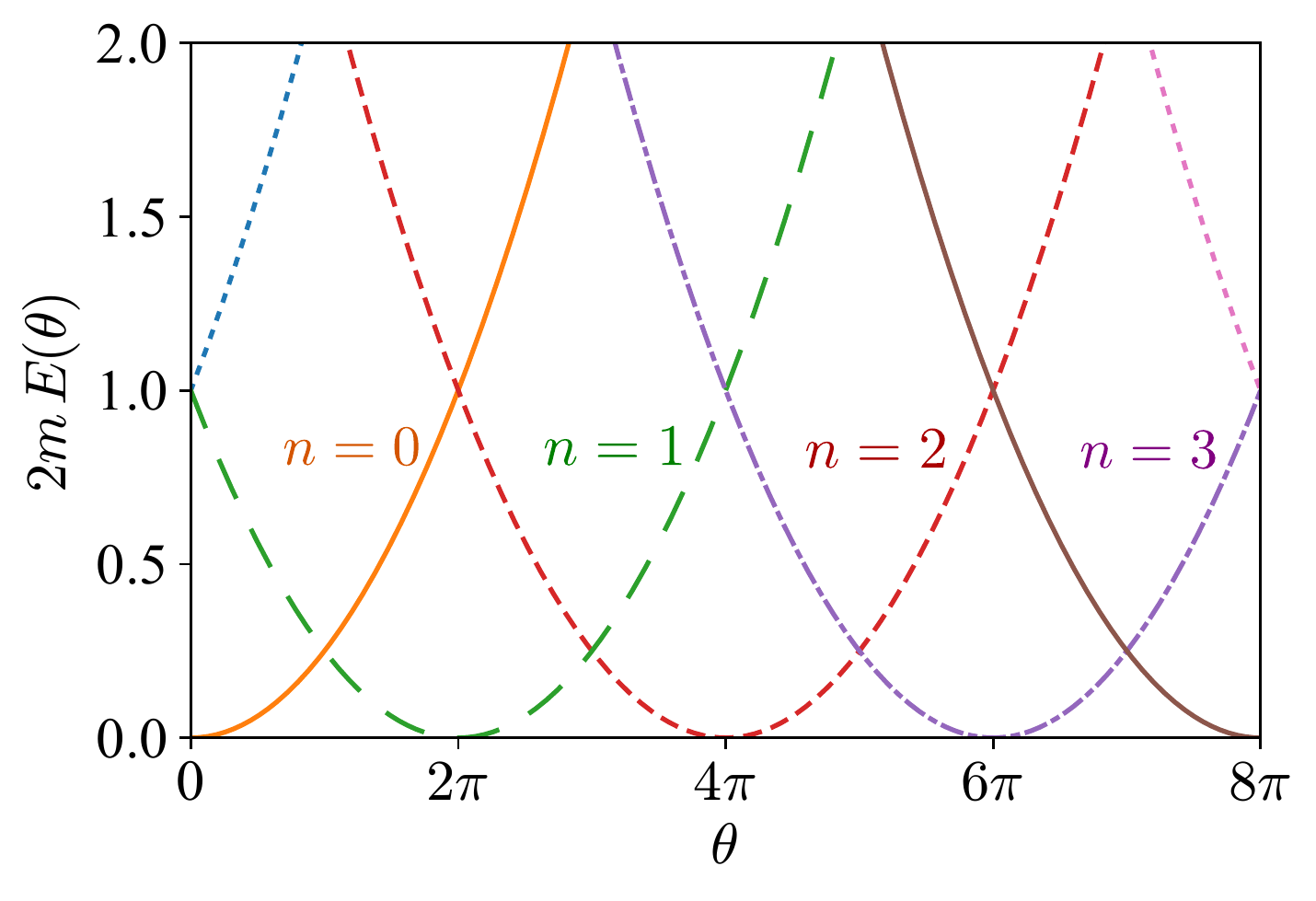}
    \caption{Eigenenergies $E_n(\theta)$ as functions of $\theta$ for various $n$'s.}
    \label{fig:Etheta}
\end{figure}

Let us consider one particle problem on the coordinate, $\phi \in S^1$, with the $2\pi$ period.
The Lagrangian takes the following form:
\begin{equation}
    L=\frac{m}{2}\dot{\phi}^2+\frac{\theta}{2\pi}\dot{\phi} - V(\phi)\,.
    \label{eq:L_S1_phi}
\end{equation}
Here, $\dot{\phi}=\diff \phi/\diff t$ denotes the derivative of $\phi$ with respect to the real time $t$ and $m$ is the mass of the particle (which can be taken to be the unity without loss of generality).
The topological $\theta$ parameter is included in the theory.
At the classical level $\theta$ is the irrelevant parameter as it does not affect the equation of motion, but nonzero $\theta$ has important physical consequences in quantum mechanics.
We can explicitly see the $\theta$ dependence in physical observables by performing the canonical quantization as explained below.
The canonical conjugate momentum is given by
\begin{equation}
    p_{\phi} = \frac{\p L}{\p \dot{\phi}}
    = m\dot{\phi} + \frac{\theta}{2\pi}\,.
\end{equation}
In this formulation the Hamiltonian, defined as $H_\theta=p_\phi\dot{\phi}-L$, has explicit dependence on the $\theta$ parameter as
\begin{equation}
    \hat{H}_\theta =
    \frac{1}{2m} \biggl( p_{\phi}
    - \frac{\theta}{2\pi} \biggr)^2
    + V(\phi)\,.
\end{equation}
In the canonical quantization, we replace $\phi\to\hat{\phi}$ and $p_\phi\to \hat{p}_\phi$ and require the canonical commutation relation, i.e., $[\hat{\phi}, \hat{p}_\phi]=\im$.
In the $\phi$-representation the canonical momentum operator is
\begin{equation}
    \hat{p}_{\phi} = \frac{1}{\im}
    \frac{\p}{\p \phi}\,.
\end{equation}
We also have to specify the Hilbert space of this system.
Because of the $S^1$ manifold with $\phi\sim \phi+2\pi$, we impose the $2\pi$ periodicity onto the wave function, i.e., $\psi(\phi+2\pi)=\psi(\phi)$.
The Hilbert space is thus spanned by
\begin{equation}
    \psi_n(\phi)={1\over \sqrt{2\pi}}\,\rme^{\im n \phi} \quad
    (n\in \mathbb{Z})\,,
\end{equation}
which constitutes an eigenstate of $\hat{p}_\phi$ with an eigenvalue $n$.
In discussions throughout this work we set $V(\phi)=0$ for simplicity.
Then, because the Hamiltonian is trivially diagonalized with the eigenstates of $\hat{p}_\phi$, we can immediately write down the eigenvalues of the Hamiltonian; that is, the eigenenergies are found to be
\begin{equation}
    E_n(\theta) = \frac{1}{2m}
    \biggl( n - \frac{\theta}{2\pi} \biggr)^2\,.
    \label{eq:eigenenergy_S1}
\end{equation}
We plot the eigenenergies as functions of $\theta$ in Figure~\ref{fig:Etheta}.
We refer to those eigenenergies for various $n$'s as the energy spectrum.

We note that the $\theta$ parameter is $2\pi$ periodic in the sense that the energy spectrum at $\theta$ should be the same as that at $\theta+2\pi$,
which is indeed the case in Figure~\ref{fig:Etheta}.
It is easy to check that the quantum system defined with $\theta+2\pi$ is unitary equivalent to the system with $\theta$, which is expressed in terms of the Hamiltonian as
\begin{equation}
    \rme^{-\im \phi}\, \hat{H}_{\theta+2\pi}\, \rme^{\im \phi} = \hat{H}_{\theta}\,.
    \label{eq:unitary_equiv}
\end{equation}
One might naively think that the $2\pi$ periodicity in $\theta$ may imply the periodicity of each eigenenergy, i.e., $E_{n}(\theta+2\pi)\stackrel{?}{=}E_{n}(\theta)$,
but this is incompatible with Eq.~\eqref{eq:eigenenergy_S1}.
The correct relation of the periodicity is
\begin{equation}
    E_{n}(\theta+2\pi)=E_{n-1}(\theta)\,,
    \label{eq:energy_relation}
\end{equation}
and this is consistent with the unitary equivalence \eqref{eq:unitary_equiv}.
This means that level crossing must occur when we continuously change $\theta$ from $0$ to $2\pi$, and the explicit formula~\eqref{eq:eigenenergy_S1} shows that it happens at $\theta=\pi$.
If we further increase $\theta$, the level crossing is located generally at $\theta=(2k-1)\pi$ $(k\in\mathbb{Z})$, as read off  from Figure~\ref{fig:Etheta}.

We can understand this level crossing behavior from the symmetry algebra.
The system has the $U(1)$ symmetry, generated by the momentum operator $\hat{p}_{\phi}$, which is more manifested in an alternative formulation in Sec.~\ref{sec:definition_deform}.
There is also the charge conjugation symmetry at $\theta=0$ or $\theta=\pi$.
The charge conjugation acts on $\phi$ and $p_\phi$ as
\begin{equation}
    \mathcal{C}: (\phi,p_{\phi})\mapsto (-\phi,-p_{\phi})\,,
\end{equation}
and this is indeed a good symmetry at $\theta=0$.
At $\theta=\pi$, however, we must modify this transformation as
\begin{equation}
    \mathcal{C}: (\phi,p_{\phi})\mapsto (-\phi,-p_{\phi}+1)\,.
\end{equation}
This shift of the momentum operator shows that the global symmetry $U(1)\rtimes (\mathbb{Z}_2)_C$ acts projectively on the Hilbert space, and thus the eigenenergies have to be doubly degenerate.
Recently, such a projective nature of symmetry is understood as an 't~Hooft anomaly for quantum mechanics, so this is a prototype of quantum field theory with nontrivial anomaly matching condition~\cite{Gaiotto:2017yup, Kikuchi:2017pcp}.

\subsection{'t~Hooft anomaly and the path integral solution}
\label{sec:pathintegral_S1}

Let us review the path integral derivation of the results obtained in the previous subsection.  This explicit derivation would be instructive for our purpose to consider the fRG approach later.
We would like to compute the thermal partition function:
\begin{equation}
    \calZ=\tr[\rme^{-\beta \hat{H}_\theta}]
    =\int \Diff \phi\,
    \exp\biggl( -\int_{0}^{\beta} \diff \tau\, L_\theta[\dot{\phi}, \phi] \biggr)\,,
\label{eq:Z_original}
\end{equation}
where the Lagrangian with the imaginary time, $\tau=\im t$, is
\begin{equation}
    L_\theta = {m\over 2}\dot{\phi}^2-\im {\theta\over 2\pi} \dot{\phi}
    + V(\phi) \,.
    \label{eq:L_original}
\end{equation}
Here, $\dot{\phi}=\diff \phi/\diff \tau=-\im \diff \phi/\diff t$.
One could have added a periodic potential, $V(\phi)$, but we will set $V(\phi)=0$ throughout this work.
As the $\theta$ term is the first-order derivative in time, it becomes pure imaginary after the Wick rotation; $t\to -\im \tau$.

For self-contained explanations, let us make a brief summary of the 't~Hooft anomaly in the present setup~\cite{Gaiotto:2017yup, Kikuchi:2017pcp}.
As we mentioned, this system has a global $U(1)$ symmetry, $\phi(\tau)\mapsto \phi(\tau)+\alpha$.
We can promote it to the local gauge redundancy by introducing a background $U(1)$ gauge field $A=A_0\diff \tau$, and the Euclidean Lagrangian with $A$ is
\begin{equation}
    L_\theta[\phi,A] = \frac{m}{2} (\dot{\phi}+A_0)^2
    - \im\frac{\theta}{2\pi} (\dot{\phi} + A_0).
\end{equation}
Remember that we set $V(\phi)=0$ for simplicity.\footnote{Even if we have a potential term like $\sim \cos(N\phi)$, we can achieve the similar conclusion by a suitable modification of the following discussion when there is a nontrivial remnant of $U(1)$ symmetry, such as $\phi\mapsto \phi+2\pi/N$. When $N$ is odd, however, we have to use the global inconsistency instead of the 't~Hooft anomaly to constrain properties of possible ground states~\cite{Gaiotto:2017yup, Kikuchi:2017pcp, Tanizaki:2017bam, Karasik:2019bxn, Tanizaki:2018xto} (see also Refs.~\cite{Cordova:2019jnf,Cordova:2019uob}).}
These two terms are obtained by the minimal coupling procedure, and they are manifestly invariant under the local transformations:
\begin{equation}
    \phi(\tau)\mapsto \phi(\tau)+\alpha(\tau), \qquad
    A_0(\tau)\mapsto A_0(\tau)-\partial_\tau \alpha(\tau),
\end{equation}
where the gauge parameter $\alpha(\tau)$ is a $2\pi$ periodic function.
The partition function in the presence of $A$ is introduced as
\begin{equation}
    \mathcal{Z}_\theta[A]=\int \Diff \phi\exp
    \biggl(-\int \diff \tau L_\theta[\phi,A]\biggr)\,.
\end{equation}
We can observe that the $2\pi$ periodicity of $\theta$ is now broken due to $A$; that is, we have
\begin{equation}
    \mathcal{Z}_{\theta+2\pi}[A]=\mathcal{Z}_\theta[A]\; \rme^{\im \int A}\,,
\end{equation}
where an additional $U(1)$ phase appears.
This $U(1)$ phase is, however, incompatible with the $\mathcal{C}$ symmetry at $\theta=\pi$.
Under $\mathcal{C}$ that transforms as
$(\phi,A) \mapsto (-\phi,-A)$, the $\theta$ angle effectively flips its sign as $\theta=\pi\mapsto -\pi$, and the partition function changes from $\calZ_{\theta=\pi}[A]$ to
\begin{equation}
    \calZ_{\theta=\pi}[\mathcal{C} A]
    = \calZ_{\theta=-\pi}[A] = \calZ_{\theta=\pi}[A]\;
    \rme^{- \im  \int A}\,.
    \label{eq:simplest_anomaly}
\end{equation}
This incompatibility of the $U(1)$ symmetry with a background gauge field and the $\mathcal{C}$ symmetry exhibits the simplest example of the mixed 't~Hooft anomaly.\footnote{We can try to eliminate the $U(1)$ anomalous phase on the right hand side of Eq.~\eqref{eq:simplest_anomaly} by adding a local counter term, and the possible choice is to multiply $\exp(-\im k\int A)$ to $\calZ_\theta[A]$.
Then, the quantization $k\in \mathbb{Z}$ is required for the large $U(1)$ gauge invariance, and it turns out that there is no suitable $k$ that can eliminate the anomalous phase. Therefore, we can conclude that Eq.~\eqref{eq:simplest_anomaly} is a genuine 't~Hooft anomaly. }
The important conclusion from the 't~Hooft anomaly is that the ground state cannot be unique, symmetric, and gapped state,
and thus we must have spontaneous breaking of the $\mathcal{C}$ symmetry.
Indeed, as seen in Figure~\ref{fig:Etheta}, the ground state is doubly degenerated at $\theta=\pi$, which reflects the presence of the mixed 't~Hooft anomaly there.

Now, let us compute $\calZ[A=0]$ to find the energy spectrum in the path-integral formulation.
Because of the gauge identification, $\phi\sim \phi+2\pi$, the periodic boundary condition for the path integral should be imposed up to this identification, that is,
\begin{equation}
    \phi(\beta)=\phi(0)+2\pi w
    \label{eq:winding_w}
\end{equation}
with $w\in \mathbb{Z}$. This integer $w$ represents the winding number, i.e., the second term in the action involving the $\theta$ angle reads:
\begin{equation}
    \int_0^\beta \diff\tau\, \frac{\theta}{2\pi} \dot{\phi}
    =\frac{\theta}{2\pi} \int \diff \phi = \theta w\,.
\end{equation}
For each topological sector characterized by $w$, we decompose the field $\phi(\tau)$ as
\begin{equation}
    \phi(\tau)={2\pi w\over \beta}\tau+\delta \phi(\tau)\,,
    \label{eq:decomposition}
\end{equation}
where $\delta\phi(\beta)=\delta\phi(0)$. We then arrive at the following expression:
\begin{equation}
    \calZ=\calN(\beta)\sum_{w=-\infty}^{\infty}\exp\biggl(-\frac{2\pi^2 m}{\beta} w^2+\im \theta w\biggr)\,.
    \label{eq:bf_Poisson}
\end{equation}
Here, the overall normalization factor, $\calN(\beta)$, comes from the path integral over $\delta \phi(\tau)$, which is independent of $w$.
Using the Poisson summation formula, we can rewrite this expression as\footnote{Here, we have chosen a suitable normalization factor $\calN(\beta)=\sqrt{2\pi m/\beta}$ to obtain the canonical expression. We can determine this factor from the semi-group property of the Feynman-Kac kernel.
}
\begin{equation}
    \calZ=\sum_{n=-\infty}^{\infty}\exp\biggl[-\frac{\beta}{2m}\biggl(n-{\theta\over 2\pi}\biggr)^2\biggr]\,.
    \label{eq:af_Poisson}
\end{equation}
In comparison with $\calZ=\sum \rme^{-\beta E_n}$, the above expression reproduces the eigenenergies in Eq.~\eqref{eq:eigenenergy_S1} obtained in the canonical quantization.
It is important to note that the quantum number $n$ in the path-integral derivation does not correspond to the winding number as one might have naively thought, but $n$ comes out as a dual variable of the winding number $w$.

\subsection{Quantum effective action and a subtle remnant of the sign problem}
\label{sec:effective_action_S1}

Here, we define the quantum effective action for this system and discuss its properties.
For this purpose, we first define the Schwinger generating functional $\mathcal{W}[J,J^*]$ for the theory represented in terms of $z=\rme^{\im\phi}\in U(1)$.
The Euclidean Lagrangian density reads:
\begin{equation}
    L_\theta = \frac{m}{2} \dot{z}^* \dot{z}
    - \frac{\theta}{4\pi} (z^* \dot{z} - \dot{z}^* z)\,,
    \label{eq:LE}
\end{equation}
where, using $z^*=z^{-1}$, we can confirm that the integration of the second term takes an integer quantized value as
\begin{equation}
    \int_0^\beta \diff\tau\, (z^* \dot{z} - \dot{z}^* z)
    = 2 \int \diff\ln z = 4\pi\im w
\end{equation}
under the boundary condition~\eqref{eq:winding_w}.
The generating functional is then given by
\begin{equation}
    \mathcal{W}[J,J^*]= \ln\int \Diff z\, \exp\left(-\int_{0}^{\beta} \diff \tau\left[{m\over 2}\dot{z}^* \dot{z}-{\theta\over 4\pi} (z^* \dot{z} - \dot{z}^* z) \right]+ z\cdot J + z^* \cdot J^* \right)\,.
    \label{eq:WJJ}
\end{equation}
Here, we introduced source fields, $J$ and $J^*$, for $z=\rme^{\im\phi}$ and $z^*=\rme^{-\im\phi}$, respectively.
Here, we adopted a short-hand notation;
$z\cdot J=\int\diff \tau z(\tau)J(\tau)$.
We also note that $\Diff z=\Diff \phi$ in the above integration is the group integration involving the Haar measure.

For $\theta=0$ this $\mathcal{W}[J,J^*]$ is a convex functional, so that we can perform the Legendre transformation to define the quantum effective action $\Gamma[Z,Z^*]$ in a usual procedure as
\begin{equation}
    \Gamma[Z,Z^*]=Z\cdot J+Z^*\cdot J^*-\mathcal{W}[J,J^*]\,.
    \label{eq:effective_action}
\end{equation}
Here, $J$ and $J^*$ on the right-hand-side of Eq.~\eqref{eq:effective_action} are implicitly determined by the following equations:
\begin{equation}
    Z(\tau)= \frac{\delta \mathcal{W}[J,J^*]}{\delta J(\tau)}\,,\qquad
    Z^*(\tau)= \frac{\delta \mathcal{W}[J,J^*]}{\delta J^*(\tau)}\,.
    \label{eq:J(Z)}
\end{equation}
It is crucial to notice that convexity of $\mathcal{W}$ ensures the uniqueness of the solution of Eq.~\eqref{eq:J(Z)}.
For nonzero $\theta\not\in 2\pi \mathbb{Z}$, as we shall see, the generating functional $\mathcal{W}[J,J^*]$ is not convex, and thus the Legendre transformation may not be well defined.
More specifically, $J$ as a function of $Z$ may not be unique from Eq.~\eqref{eq:J(Z)} without the convex property, and the definition of $\Gamma[Z,Z^*]$ in Eq.~\eqref{eq:effective_action} cannot avoid ambiguity.
Since the fRG formalism of Wetterich type treats not $\mathcal{W}[J,J^*]$ but $\Gamma[Z,Z^*]$, it is a crucial problem if $\Gamma[Z,Z^*]$ can really exist or not for general systems with the sign problem.
It is often said that the fRG does not rely on importance sampling and does not have the sign problem at all, but one should be cautious about the existence the Legendre transformation for such theories that suffer the sign problem.
In this section we demonstrate that, in the present case of the $S^1$ quantum mechanics, a concrete construction of $\Gamma[Z,Z^*]$ for any $\theta$ is possible as a formal power series in terms of $Z$ and $Z^*$ using the same formula~\eqref{eq:effective_action}.
In other words, we solve Eq.~\eqref{eq:J(Z)} in an iterative way by imposing an ansatz for $J$ and $J^*$ at $\theta\not\in 2\pi\mathbb{Z}$; that is, $J,J^*\to 0$ in the limit of $Z,Z^*\to 0$ is presumed.
We note that this ansatz is justified in the present case since quantum mechanics with finite degrees of freedom does not spontaneously break continuous symmetry.  For general problems in quantum field theories the remnant of the sign problem must be treated carefully.

It might sound strange that we must put an extra assumption on the vacuum to define the quantum effective action, as we usually search for the quantum vacuum by finding the minimum of $\Gamma$, and the logic here seems to go the other way around.
We can make our point more explicit by writing down several lines of expressions.  To see a potential failure in the standard procedure, let us revisit the convexity of $\mathcal{W}[J,J^*]$.
Assuming smoothness, $\mathcal{W}[J,J^*]$ is convex iff
\begin{equation}
    \left(\rho\cdot{\delta\over \delta J}+\rho^*\cdot {\delta\over \delta J^*}\right)^2 \mathcal{W}[J,J^*]\ge 0
    \label{eq:convex}
\end{equation}
for any $\rho(\tau)\in \mathbb{C}$.
We can rewrite this condition as
\begin{equation}
    \left\langle \left[\rho\cdot(z-\langle z\rangle_J)+\rho^*\cdot (z^*-\langle z^*\rangle_J)\right]^2 \right\rangle_J \ge 0\,,
    \label{eq:lhs_convex}
\end{equation}
where $\langle O[z,z^*] \rangle_J= e^{-W} \int \Diff z\, O[z,z^*] \exp(-S+ z\cdot J+z^*\cdot J^*)$.
When the $\theta$ term is absent, i.e., $\theta=0$, the Euclidean action $S$ is real, and thus the integrand of Eq.~\eqref{eq:lhs_convex} is positive semi-definite, which proves Eq.~\eqref{eq:convex}.
For $\theta\neq 2\pi\mathbb{Z}$, the Euclidean action $S$ takes a complex value that causes the sign problem, and the positivity condition~\eqref{eq:convex} may be violated.

We can furthermore check that the condition \eqref{eq:convex} holds only if $\theta\in 2\pi \mathbb{Z}$ by putting $J=J^*=0$.
When we set $J=J^*=0$, we immediately see that $\langle z\rangle=\langle \rme^{\im \phi}\rangle=0$ and $\langle z(\tau_1)z(\tau_2)\rangle=\langle \rme^{\im \phi(\tau_1)}\rme^{\im \phi(\tau_2)}\rangle=0$ because of  $U(1)$ symmetry.
Therefore, Eq.~\eqref{eq:convex} at $J=J^*=0$ turns out to be equivalent to
\begin{equation}
    \int \diff \tau_1 \diff \tau_2\, \rho(\tau_1)\rho^*(\tau_2) G(\tau_1-\tau_2)\ge 0\,,
    \label{eq:convex_g}
\end{equation}
where we have introduced the following two-point function:
\begin{equation}
    G(\tau_1-\tau_2)=
    \langle z(\tau_1) z^*(\tau_2)\rangle_{J=0} = \langle \rme^{\im \phi(\tau_1)} \rme^{-\im\phi(\tau_2)}\rangle_{J=0}\,.
\end{equation}
A quick calculation shows that Eq.~\eqref{eq:convex_g} is satisfied for any $\rho$ and $\rho^*$ iff
\begin{equation}
    G(\tau)\ge 0,\qquad G(\tau)=G(-\tau)\,.
\end{equation}
Let us explicitly compute $G(\tau)$ for $-\pi<\theta<\pi$ at $\beta\to \infty$.
In this circumstance only the ground state $|0\rangle$ with its energy $E_0$ contributes to the partition function, which simplifies the computation of $G(\tau)$ in the canonical quantization:
\begin{align}
G(\tau)&= \langle 0| \left[\Theta(\tau) \rme^{\im \phi}\exp(-\tau(\hat{H}-E_0(\theta)))\rme^{-\im \phi}
+\Theta(-\tau)\rme^{-\im \phi}\exp(\tau(\hat{H}-E_0(\theta)))\rme^{\im\phi}\right]|0\rangle\notag\\
&= \Theta(\tau)\exp(-\tau(E_{-1}(\theta)-E_0(\theta)))+\Theta(-\tau)\exp(\tau (E_1(\theta)-E_0(\theta)))\,,
\label{eq:G(tau)}
\end{align}
where the previous relations~\eqref{eq:unitary_equiv} and \eqref{eq:energy_relation} are used from the first to the second line.
The above expression shows that the first condition, $G(\tau)\ge 0$, is satisfied, while the second condition, $G(\tau)=G(-\tau)$, requires $E_1(\theta)=E_{-1}(\theta)$.
From the explicit form~\eqref{eq:eigenenergy_S1} $E_1(\theta)=E_{-1}(\theta)$ is true only if $\theta=0$.  Thus, the convexity of $\mathcal{W}[J,J^*]$ is verified for $\theta=0$ but it is violated for nonzero $\theta$.
In other words, Eq.~\eqref{eq:J(Z)} may not have a unique solution without extra assumption.

Now, let us argue that we can solve Eq.~\eqref{eq:J(Z)} for $J$, $J^*$ in terms of $Z$, $Z^*$ as a formal power series around $J=J^*=0$.
Because of the $U(1)$ symmetry\footnote{Without the $U(1)$ symmetry, the following discussion does not hold, so we must revise the analysis for a general case with symmetry breaking potential terms such as $V(\phi)\sim \cos(N\phi)$.}, $Z|_{J=0}=\langle \rme^{\im\phi}\rangle_{J=0}=0$, and thus
\begin{align}
    Z(\tau) &= \int\diff \tau'\,  G(\tau-\tau')J^*(\tau') + \calO(|J|^3)\,,\\
    Z^*(\tau) &= \int \diff\tau'\, J(\tau')G(\tau'-\tau) + \calO(|J|^3)\,.
\label{eq:ZGJ}
\end{align}
When $-\pi<\theta<\pi$ and $\beta\to \infty$, $G(\tau)$ is given by Eq.~\eqref{eq:G(tau)}, which satisfies:
\begin{equation}
    \left[-m \p_\tau^2-{\theta\over \pi}\p_\tau+{1\over 4m}\left(1-{\theta^2\over \pi^2}\right)\right]G(\tau)=\delta(\tau).
    \label{eq:diff_G}
\end{equation}
This gives an operator that is an inverse of $G(\tau_1-\tau_2)$, and we can solve Eq.~\eqref{eq:ZGJ} as
\begin{align}
    J(\tau) &= \left[-m\p_\tau^2+{\theta\over \pi}\p_\tau+{1\over 4m}\left(1-{\theta^2\over \pi^2}\right)\right]Z^*(\tau) + \calO(|Z|^3)\,,\\
    J^*(\tau) &= \left[-m\p_\tau^2-{\theta\over \pi}\p_\tau+{1\over 4m}\left(1-{\theta^2\over \pi^2}\right)\right]Z(\tau) + \calO(|Z|^3)\,.
\end{align}
It is now evident that we can iterate this procedure to construct $J$, $J^*$ as a formal power series of $Z$, $Z^*$.
We thus find that the quantum effective action is perturbatively expressed as
\begin{equation}
    \Gamma[Z,Z^*]=\beta E_0(\theta)+\int\diff \tau \left[ m|\dot{Z}|^2-{\theta\over 2\pi}(Z^* \dot{Z}-Z \dot{Z}^*)+{\pi^2-\theta^2\over 4\pi^2m}|Z^2|\right] + \calO(|Z^4|)
    \label{eq:Gamma_order2}
\end{equation}
for $|\theta|<\pi$ and $\beta\to\infty$.
In Sec.~\ref{sec:Villain}, we give another derivation of \eqref{eq:Gamma_order2} using the Villain-type lattice regularization.
This is already a highly nontrivial result.
We point out that the kinetic terms, $|\dot{Z}|^2$ and $(Z^*\dot{Z}-Z\dot{Z}^*)$, in the quantum effective action~\eqref{eq:Gamma_order2} are multiplied by the factor $2$ compared with the classical action~\eqref{eq:WJJ}.
Let us also emphasize again that this quantum effective action $\Gamma[Z,Z^*]$ is not convex for $\theta\not=0$.  The linear term in $\theta$ is pure imaginary, so $\Gamma[Z,Z^*]$ is a complex-valued functional for $\theta\not=0$.
Because of this complex property, we cannot identify the quantum vacuum by minimizing $\Gamma$; such a notion is no longer defined for theories with the sign problem generally.
Instead, as $\Gamma$ is defined as a formal power series, we can obtain the ground-state energy from $\Gamma$ by setting $Z=Z^*=0$.
From the quadratic term in $Z$, $Z^*$, we can obtain energies of the first and the second excited states.

Let us discuss how we can understand the microscopic structure of the level crossing and the shift of the ground-state energy at $\theta=\pi$ from the above quantum effective action~\eqref{eq:Gamma_order2}.
As we see from Figure~\ref{fig:Etheta}, $E_0(\theta)$ and $E_{1}(\theta)$ cross at $\theta=\pi$ and this phenomenon is similar to a first-order phase transition.
There, one might naively expect that this should be described by a jump of the quantum vacuum from $Z=Z^*=0$ to another location. However, this is not the case, because the $U(1)$ symmetry is unbroken on both sides.
As $Z=Z^*=0$ is the unique point that is symmetric under $U(1)$, the quantum vacuum must be at $Z=Z^*=0$ for both $\theta<\pi$ and $\theta>\pi$.
Indeed, repeating the same computations for shifted $\theta$, i.e., $-\pi<\theta-2\pi<\pi$, we find the identical form of the quantum effective action \eqref{eq:Gamma_order2} with $\theta$ replaced by $\theta-2\pi$.
We note that $\theta^2$ and $(\theta-2\pi)^2$ are continuous but has a cusp at $\theta=\pi$, and the coefficient of $Z^*\dot{Z}-Z \dot{Z}^*$ exhibits a discrete jump from $\pi$ to $-\pi$ when crossing $\theta=\pi$.
These serve as signals for the phase transition, but we have to treat non-analytic features of effective actions to observe them.

Lastly, let us summarize several remarks on general lessons we should learn about the technical aspect.
The lack of convexity of the generating functional $\mathcal{W}$ is a generic phenomenon when the path integral suffers from the sign problem.
In such a situation with the sign problem, we cannot uniquely define the quantum effective action $\Gamma$ as the Legendre transform of $\mathcal{W}$ without extra prescription.
A possible detour may be to use not $\Gamma$ but the generating functional $\mathcal{W}$ itself.  Alternatively, the constrained effective potential~\cite{Fukuda:1974ey} instead of the 1PI effective action $\Gamma$ would be useful.
When we take the constrained effective potential approach, the reincarnation of the sign problem appears as the absence of the saddle point in terms of the original path-integral variables~\cite{Fukushima:2006uv}, and the complex saddle points become important for the quantum vacuum~\cite{Tanizaki:2015pua,Tanizaki:2016xcu}.
For the present quantum mechanical system, thanks to the unbroken $U(1)$ symmetry, we do not have to worry too much about this issue of the sign problem.  In the rest of this paper the quantum effective action $\Gamma$ should be always understood as the one constructed by a formal series in $Z$, $Z^*$.

\section{Relaxing the topology in the fRG approach}

The fRG equation (e.g., the Wetterich equation) provides us with a quantization scheme to calculate the full quantum effective action.
Since the fRG is formulated in a differential form, however, the topological $\theta$ term is entirely dropped as long as the topological number $w$ is an integer.
To evade this problem, in Sec.~\ref{sec:definition_deform}, we propose a deformation of the theory to smear the $S^1$ winding with auxiliary parameter $g$.
In Sec.~\ref{sec:canonical_deform} we write down the Schr\"{o}dinger equation for the deformed theory and numerically solve the eigenenergies as functions of $\theta$ and $g$.
Then we test a common approximation scheme in the fRG approach to find that the dependence of the ground-state energy on $\theta$ and $g$ can be partially reproduced.

\subsection{Deforming the theory to unquantize the winding number}
\label{sec:definition_deform}

We shall deform the theory originally defined by the Lagrangian~\eqref{eq:LE}.
We relax the condition to quantize the integer winding by changing the variable as
\begin{equation}
    z=\rme^{\im \phi} ~~\longrightarrow~~ u=r \rme^{\im q}\,,
\end{equation}
for which $u^*\neq u^{-1}$.
This means that $u\not\in U(1)$, and thus the $\pi_1(U(1))$ winding is explicitly broken.
Accordingly, we propose a deformed theory as follows:
\begin{align}
    L_\theta(z) ~ \longrightarrow ~ L_\theta(u) &= \frac{m}{2} \dot{u}^* \dot{u}
    - \frac{\theta}{4\pi}
    (u^*\dot{u}-\dot{u}^* u)
    + \frac{g}{4}(u^* u-1)^2 \notag\\
    &= \frac{m}{2} (\dot{r}^2 + r^2 \dot{q}^2)
    - \im \frac{\theta}{2\pi} r^2 \dot{q}
    + \frac{g}{4}(r^2 - 1)^2\,.
    \label{eq:L_deform}
\end{align}
In this theory we included a wine-bottle potential $\propto g$ so that we can extrapolate the deformed theory to the original one.
Actually, in the limit of $g\to\infty$, the potential term constraints the theory to be on $r=1$ only.  Then, $u$ with $r=1$ is reduced to $z\in U(1)$ and the original theory is recovered.

\begin{figure}
    \centering
    \includegraphics[width=0.5\textwidth]{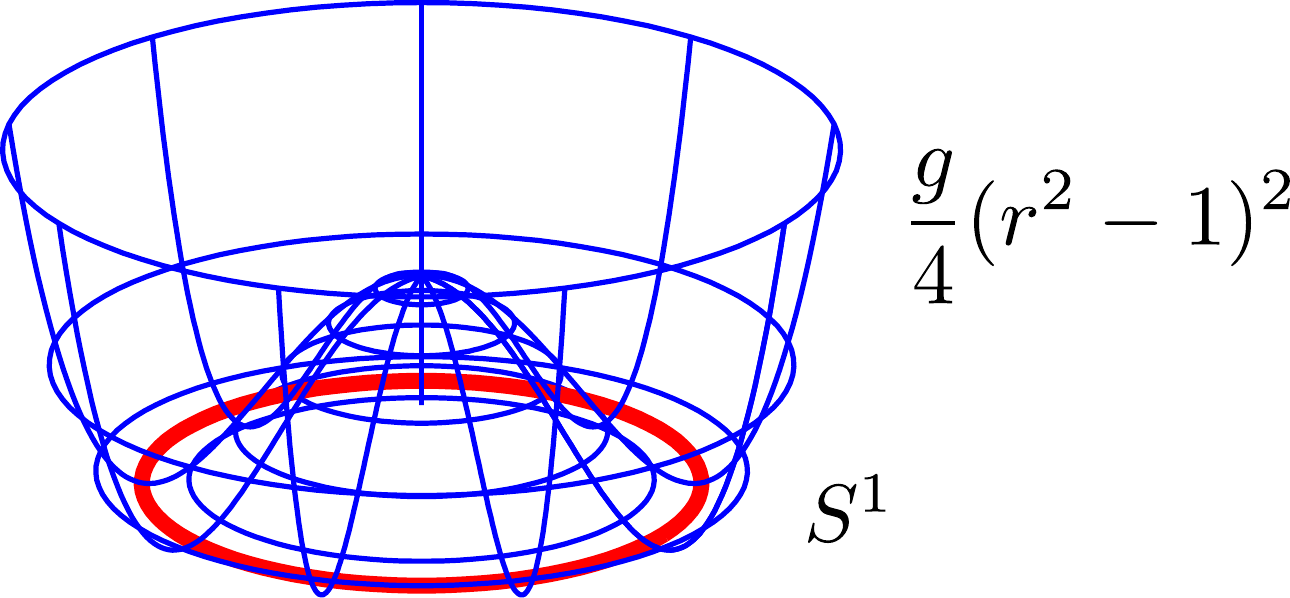}
    \caption{Potential in the deformed theory.  The original theory on $S^1$ lies at the bottom of the potential, and the original theory is recovered in the $g\to\infty$ limit.}
    \label{fig:deformed}
\end{figure}

Roughly speaking, the idea is that the theory space is augmented with an extra parameter $g$, so that the coefficient in the $\theta$ term is no longer an integer for $g<\infty$ and the $\theta$ dependence can emerge also in the fRG formalism.
We cannot find the analytical solution for general $g$, but it is easy to evaluate the path integral at $g=0$.  Although we want to take the $g\to\infty$ limit in the end, it is instructive to go through the analytical integration at $g=0$ first.

The thermal partition function at $g=0$ reads:
\begin{equation}
    \calZ_{0} = \int\Diff u \Diff u^*\,
    \exp\left[-\int_0^\beta \diff\tau \left(
    \frac{m}{2}\dot{u}^* \dot{u} - \frac{\theta}{4\pi}
    (u^* \dot{u} - \dot{u}^* u) \right) \right]\,.
\end{equation}
We can take the Matsubara sum in the standard procedure and we eventually find,
\begin{equation}
    \calZ_{0} = \infty\times \sum_{n=0}^\infty
    \exp\left[-\frac{\beta|\theta|}{\pi m}
    \left(n+\frac{1}{2}\right) \right],
\end{equation}
and there is an overall divergent constant.
This divergence is special for $g=0$, and the partition function becomes finite for $g>0$.
To have the physical interpretation of this result, we note that this system at $g=0$ is equivalent to the quantum mechanics on the plane with the uniform magnetic field by choosing the symmetric gauge.
Each energy eigenvalue corresponds to the Landau level,
\begin{equation}
    E_n(\theta;g=0) = \frac{|\theta|}{\pi m}
    \left( n+\frac{1}{2} \right)\,,
\end{equation}
and the divergent constant represents the infinite degeneracy of each Landau level.

This expression looks totally different from Eq.~\eqref{eq:eigenenergy_S1} at $g\to\infty$.
As a benchmark for the fRG approach, it is desirable to know $E_n(\theta;g)$ for general $0<g<\infty$.
We find that the numerical calculations turn out to be straightforward in the canonical formalism rather than the path-integral approach.

\subsection{Numerical solutions in the canonical quantization}
\label{sec:canonical_deform}

For the canonical quantization we should construct the Hamiltonian from the theory definition by Eq.~\eqref{eq:L_deform}.
For convenience in the canonical quantization we convert Eq.~\eqref{eq:L_deform} to the real-time convention for which $-\im$ in the coefficient of the $\theta$-dependent term in Eq.~\eqref{eq:L_deform} turns to be the unity.
The canonical momenta for $r$ and $q$ are, respectively,
\begin{equation}
    \hat{\Pi}_r = m\dot{r}\,,\qquad
    \hat{\Pi}_q = mr^2\dot{q} + \frac{\theta}{2\pi}r^2\,.
\end{equation}
The Hamiltonian is thus given by
\begin{equation}
    \hat{H}_g = \frac{1}{2m}\hat{\Pi}_r^2
    + \frac{1}{2mr^2}\left(\hat{\Pi}_q - \frac{\theta}{2\pi}r^2 \right)^2
    + \frac{g}{4}(r^2-1)^2\,.
\end{equation}
Let us adopt a unit system with $m=1$ for notational brevity.  It is easy to restore the full $m$ dependence if necessary.
Then, in the coordinate representation, the canonical momenta are written in terms of the derivative operators:
\begin{equation}
    \hat{\Pi}_r^2 = -\frac{1}{r}\frac{\partial}{\partial r} \Bigl(r \frac{\partial}{\partial r} \ \Bigr)\,,\qquad
    \hat{\Pi}_q = -\im \frac{\partial}{\partial q}\,.
\end{equation}
Once the angular dependence is mode expanded with the plane-wave basis, $\rme^{\im n q}$ $(n\in \mathbb{Z})$, then $\hat{\Pi}_q$ can be simply replaced with $n$.  Then, the Schr\"{o}dinger equation is
\begin{equation}
    \biggl[ -\frac{1}{2r}\frac{\partial}{\partial r} \Bigl( r\frac{\partial}{\partial r} \Bigr) + \frac{1}{2r^2} \Bigl( n - \frac{\theta}{2\pi}r^2 \Bigr)^2 + \frac{g}{4}(r^2-1)^2 \biggr]\psi_{\ell,n}(r)
    = E_{\ell,n}(\theta;g)\,\psi_{\ell,n}(r)\,.
    \label{eq:Emn}
\end{equation}
We can solve the above bound-state problem with the boundary condition, $\psi_{\ell,n}(r\to\infty)\to 0$, to find discretized $E_{\ell,n}$.  Here, we note that $\ell$ represents the quantum number associated with the radial excitation, and we are interested in the ground state energy at $\ell=0$.
Hereafter we suppress the subscript $\ell$ by taking $\ell=0$ only: $E_n(\theta;g)=E_{\ell=0,n}(\theta;g)$.

\begin{figure}
    \centering
    \includegraphics[width=0.7\textwidth]{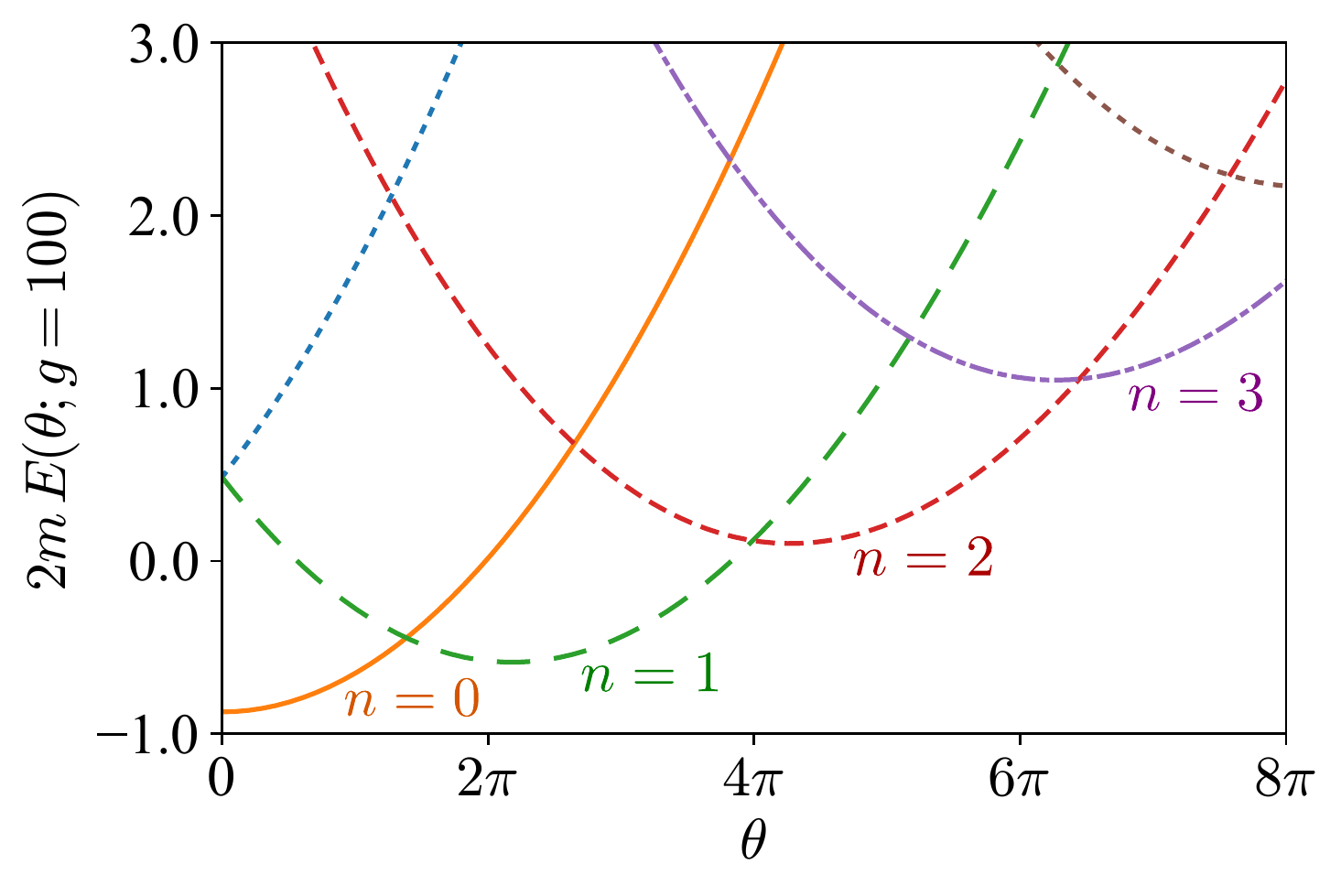}
    \caption{Eigenenergies $E_n(\theta;g)$ at $g=100$ for various $n$ as functions of $\theta$.
    The energy is shifted by an offset of $-\sqrt{g/2}\simeq -7.1$.}
    \label{fig:Etheta_g}
\end{figure}

The eigenenergies, $E_n(\theta;g)$, cannot be directly compared to $E_n(\theta)$ we obtained previously.
This is because the energy is lifted up by a potential $\propto g(r^2-1)^2$ in which nontrivial $g$ dependence may remain even in the $g\to \infty$ limit due to the zero-point energy.
We can estimate this remaining $g$ dependence at large $g$ as follows. Let us expand the Schr\"{o}dinger equation as $r=1+\rho$ with a shifted eigenenergy,
$E_n(\theta) + \delta E(g)$, as
\begin{equation}
  \Bigl[ -\frac{1}{2} (\partial_\rho^2 + \partial_\rho) + g \rho^2 \Bigr] \chi(\rho) = \delta E(g) \chi(\rho)
\end{equation}
This is an eigenequation for a harmonic oscillator whose ground state wave-function is
\begin{equation}
    \chi(\rho) = \exp\biggl(-\sqrt{\frac{g}{2}}\rho^2 - \frac{1}{2}\rho \biggr)
\end{equation}
up to an irrelevant normalization factor whose eigenenergy is immediately found to be
\begin{equation}
    \delta E(g) = \sqrt{\frac{g}{2}}\,.
    \label{eq:energy_sub}
\end{equation}
Therefore, we must subtract this $g$ dependent term to recover $E_n(\theta)$ in the $g\to\infty$ limit:
\begin{equation}
    E_n(\theta;g)-\sqrt{g\over 2}=\mathrm{const.}+{1\over 2}\left(n-{\theta\over 2\pi}\right)^2+\mathcal{O}(1/\sqrt{g}).
\end{equation}
Figure~\ref{fig:Etheta_g} shows the eigenenergies, $E_n(\theta;g)$, obtained numerically from Eq.~\eqref{eq:Emn} with a subtraction by Eq.~\eqref{eq:energy_sub}.
We see that the qualitative behavior is similar to the eigenenergies in the original theory as shown in Figure~\ref{fig:Etheta}.

\subsection{Local potential approximation in the fRG approach}
\label{sec:fRG_deform}

It is intriguing to apply the fRG equation to the deformed theory defined in Eq.~\eqref{eq:L_deform}.
As an etude, let us first consider an even simpler problem of the two-dimensional harmonic oscillator for which we know the exact answer.
This might sound a trivial check, but is a meaningful detour;  how to extract the ground state energy in the fRG framework is not so trivial.

The quantum effective action in the LPA assumes:
\begin{equation}
    \Gamma_k = \int \diff\tau \biggl[
    \frac{m}{2}\dot{u}^* \dot{u} + V_k(|u|^2)\biggr]\,.
    \label{eq:harmonic_LPA}
\end{equation}
The complex variable is reparametrized as
$u=x+\im y$ and $r^2 = |u|^2 = x^2 + y^2$.  Then, the initial potential for the harmonic oscillator problem at the scale $k=\Lambda$ should be
\begin{equation}
    V_\Lambda = \frac{1}{2}m\omega^2 r^2\,.
\end{equation}
We note that $\Lambda$ is a UV scale from which the renormalization group flow is started.
The Wetterich equation reads:
\begin{equation}
    \partial_k \Gamma_k =
    \frac{1}{2}\,
    \mathrm{Tr} \biggl[ \partial_k R_k
    \biggl( \Gamma_k^{(2)} + R_k
    \biggr)^{-1} \biggr]\,.
    \label{eq:Wetterich}
\end{equation}
In the basis of Fourier transformed $\tilde{x}(p)$ and $\tilde{y}(p)$, we can compute a matrix as
\begin{equation}
    \Gamma_k^{(2)} =
    \begin{pmatrix}
        \displaystyle \frac{\delta^2 \Gamma_k}{\delta\tilde{x}(-p)\delta\tilde{x}(p)} & \displaystyle \frac{\delta^2 \Gamma_k}{\delta\tilde{x}(-p)\delta\tilde{y}(p)} \\[12pt]
        \displaystyle \frac{\delta^2 \Gamma_k}{\delta\tilde{y}(-p)\delta x(p)} & \displaystyle \frac{\delta^2 \Gamma_k}{\delta\tilde{y}(-p)\delta\tilde{y}(p)}
    \end{pmatrix}
    =
    \begin{pmatrix}
        mp^2 + V_{xx} ~&~ \displaystyle V_{xy} \\[12pt]
        \displaystyle V_{xy} ~&~ mp^2 + V_{yy}
    \end{pmatrix}\,,
\end{equation}
where we can express the derivatives as
\begin{equation}
    V_{xx} = 4x^2 V'' + 2V' \,,\qquad
    V_{yy} = 4y^2 V'' + 2V' \,,\qquad
    V_{xy} = 4xy V''
\end{equation}
with $V'$ and $V''$ representing the first and the second derivatives in terms of $r^2$.
It is straightforward to take the inverse of the above matrix and take the trace.  After all, we find the following equation:
\begin{equation}
    \partial_k V_k = \int\frac{dp}{2\pi}
    \frac{\partial_k R_k\, (mp^2 + 2V' + 2r^2 V'' + R_k)}{\displaystyle (mp^2+2V'+2r^2 V''+R_k)^2 - (2r^2 {V''})^2}\,.
\end{equation}
It is a common technique to use Litim's optimized regulator, i.e.,
\begin{equation}
    R_k(p) = m(k^2 - p^2)\Theta(k^2 - p^2)
\end{equation}
with the Heaviside step function~\cite{Litim:2001up}.  This convenient choice of the regulator eliminates the $p$ dependence and then the $p$-integration amounts to the phase-space volume.  That is, the fRG equation in the LPA is
\begin{equation}
    \partial_k V_k = \frac{2mk^2}{\pi}
    \frac{mk^2 + 2V' + 2r^2 V''}
    {(mk^2 + 2V' + 2r^2 V'')^2 - (2r^2 {V''})^2}\,.
\end{equation}
In the case of the harmonic oscillator, we already know that the quantum fluctuations would never produce higher-order than quadratic terms.
Therefore, we can safely fix $V'=\tfrac{1}{2}m\omega^2$ and $V''=0$, so that we can simplify the fRG equation and easily perform the $k$-integration as
\begin{equation}
    V_0 = V_\Lambda + \int_\Lambda^0 \diff k\, \frac{2mk^2}{\pi (mk^2 + m\omega^2)}
    = \frac{1}{2}m\omega^2\, r^2
    + \frac{2\omega}{\pi} \arctan(\Lambda/\omega) - \frac{2}{\pi} \Lambda\,.
\end{equation}
We see that, in the limit of $\Lambda/\omega\to\infty$, the second term correctly reproduces the ground state energy; that is, twice of the zero-point energy, $\omega/2$, amounts to $\omega$.
An important lesson we can learn is that the convergence to the correct value is, however, logarithmically slow.
For example, $(2/\pi)\arctan(10)\simeq 0.937$, which means that the deviation remains more than $6\%$ even for $\Lambda$ as large as $10\omega$.
Another useful observation is the appearance of the last term proportional to $\Lambda$.
Thus, such a UV divergent term should be subtracted.
We can simply make the subtraction not after the integration but already in the integrand by a term set with $V'=V''=0$.
In the above example of the harmonic oscillator, this subtraction leads to the regularized integrand as
\begin{equation}
    \frac{2mk^2}{\pi(mk^2 + m\omega^2)} ~\to~
    \frac{2mk^2}{\pi(mk^2 + m\omega^2)}
    - \frac{2mk^2}{\pi mk^2}
    = \frac{2m\omega^2}{\pi(mk^2 + m\omega^2)}\,.
\end{equation}
from which the finite part is directly derived.

Now, we are well armed with calculations in the analytically solvable example, and we shall proceed to the LPA application to our deformed theory.
In the LPA treatment we employ the following form of the scale-dependent effective action:
\begin{equation}
    \Gamma_k = \int \diff\tau \biggl[
    \frac{m}{2}\dot{u}^* \dot{u}
    -\frac{\theta}{4\pi} (u^* \dot{u} - \dot{u}^* u)
    + V_k(|u|^2) \biggr]\,.
    \label{eq:Gamma_LPA}
\end{equation}
This is a straightforward extension of Eq.~\eqref{eq:harmonic_LPA} with the $\theta$ term.  The initial potential is changed into
\begin{equation}
    V_\Lambda = \frac{g}{4} (r^2 -1)^2\,.
    \label{eq:VLambda}
\end{equation}
The Wetterich equation itself is common, but the matrix elements are slightly modified by the $\theta$ term, i.e.,
\begin{equation}
    \Gamma_k^{(2)} =
    \begin{pmatrix}
        mp^2 + V_{xx} ~&~ \displaystyle -\frac{\theta p}{\pi} + V_{xy} \\[12pt]
        \displaystyle \frac{\theta p}{\pi} + V_{xy} ~&~ mp^2 + V_{yy}
    \end{pmatrix}\,.
\end{equation}
In the same way, as we did in the harmonic oscillator problem, we can take the inverse of this matrix with the choice of Litim's optimized regulator, eventually reaching:
\begin{equation}
    \partial_k V_k = \frac{mk}{\pi}
    \biggl(mk^2 + 2V' + 2r^2 V''\biggr)
    \int_{-k}^k \frac{dp}
    {\displaystyle \frac{\theta^2}{\pi^2} p^2 + (mk^2 + 2V' + 2r^2 V'')^2 - (2r^2 {V''})^2}\,.
\end{equation}
A crucial difference from the harmonic oscillator problem is that the $p$-integration is not a mere phase-space volume, but the denominator has a $p$-dependent term.
If we assume $mk^2 + 2V'\ge 0$ (that is not always the case as we will see later), we can immediately perform the $p$-integration to obtain:
\begin{equation}
    \partial_k V_k
    = \frac{2mk}{|\theta|}
    \frac{mk^2 + 2V' + 2r^2 V''}
    {\sqrt{(mk^2 + 2V' + 2r^2 V'')^2 - (2r^2 V'')^2}} \,\phi(k,\theta)
    - \frac{2}{\pi}\,,
    \label{eq:fRG_LPA}
\end{equation}
where we introduced an angle variable given by
\begin{equation}
    \phi(k,\theta) = \arctan\Biggl(
    \frac{k|\theta|}{\pi\sqrt{(mk^2+2V'+2r^2 V'')^2 - (2r^2 V'')^2}}\Biggr)\,.
\end{equation}
The last term, $-2/\pi$, is the subtraction to eliminate the UV divergence.

\begin{figure}
    \centering
    \includegraphics[width=0.7\textwidth]{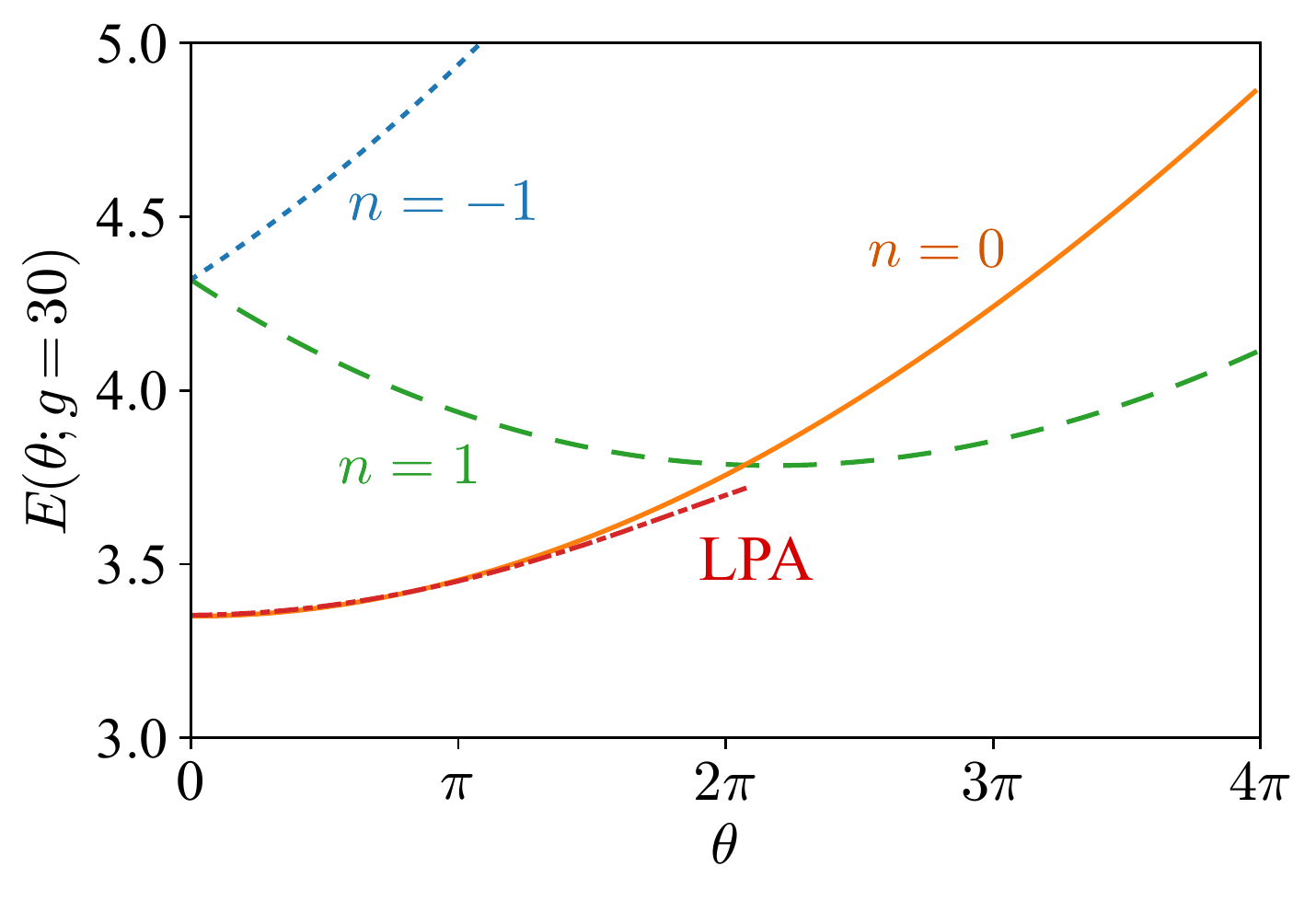}
    \caption{Eigenenergies $E_n(\theta;g)$ at $g=30$ as functions of $\theta$.  The dot-dashed curve represents the fRG results in the LPA\@.
    The energy is not shifted here.} \label{fig:ELPA}
\end{figure}

One may think that it is a straightforward task to perform the $k$-integration numerically and find the ground state energy from $V_{k\to 0}$.
There are, however, two extremely nontrivial features in the numerical calculations.
Before explaining the numerical setup in details, let us discuss such nontrivial behavior.

\begin{figure}
    \centering
    \includegraphics[width=0.7\textwidth]{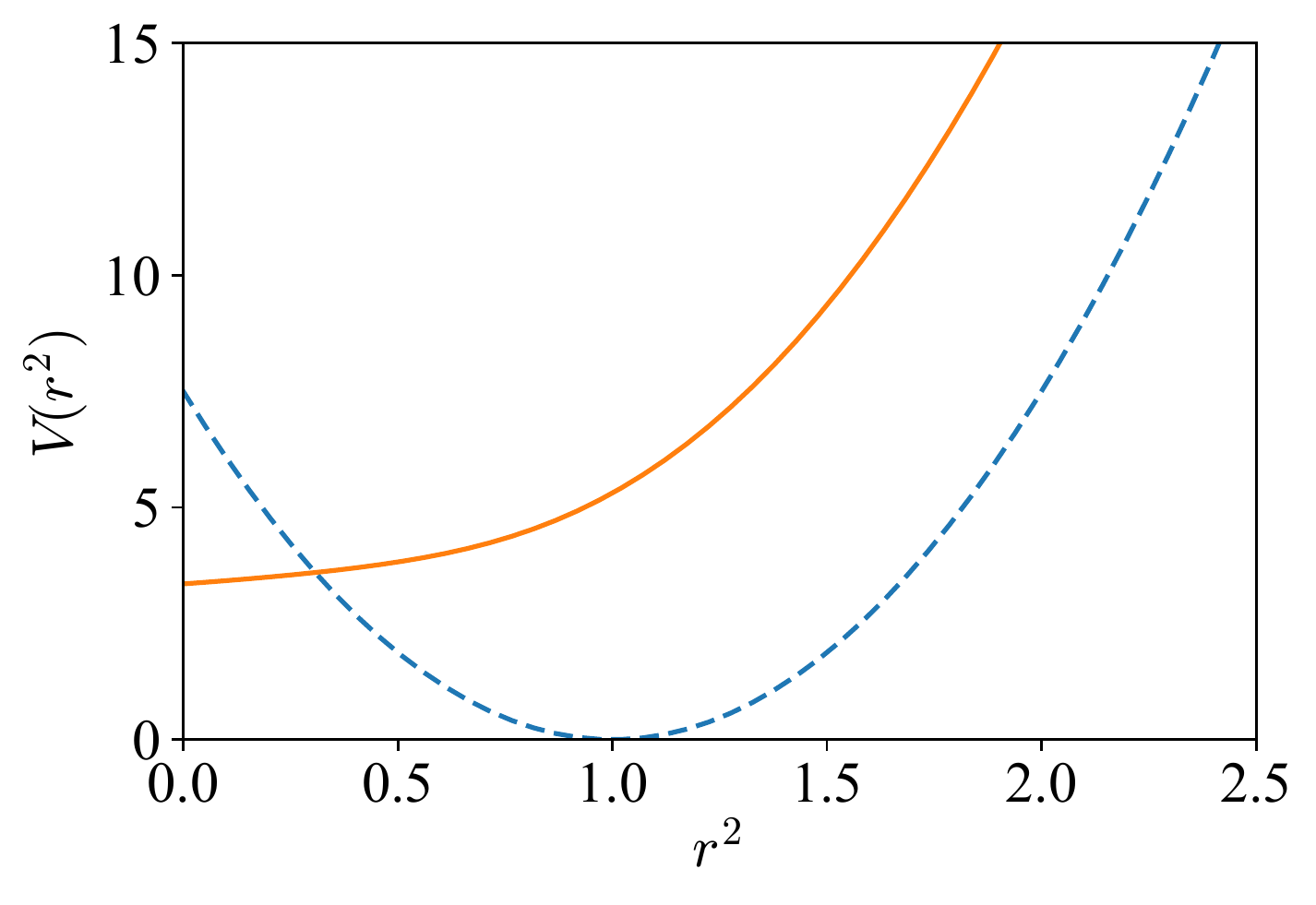}
    \caption{Effective potential at $\theta=0$ evolved from $k=\Lambda$ to $k=10^{-3}$ (where $\Lambda=3000$).  The dotted curve represents the initial form of the effective potential, $V_\Lambda$, in Eq.~\eqref{eq:VLambda} with $g=30$.} \label{fig:poten_t00}
\end{figure}

The first one is that the minimum of the effective potential is inevitably located at $r=0$.  To visualize the potential shape with quantum fluctuations integrated out, we plot the effective potential at $\theta=0$ in Figure~\ref{fig:poten_t00}.  The dotted curve represents the tree-level potential at $k=\Lambda$ that has a global minimum at $r=1$.  We note that the original theory is recovered in the $g\to \infty$ limit in which $u=r \rme^{\im \theta}$ is subject to be an element of $U(1)$ with $r=1$ fixed.  As we have verified in the canonical quantization method in the previous subsection, the energy spectrum from the modified theory certainly approaches the one from the original theory.  At the quantum level, however, the effective potential should be convex in general, and moreover, spontaneous symmetry breaking is not possible in quantum mechanics where physical degrees of freedom are not infinite.  Therefore, the full quantum effective potential must have a global minimum only at $r=0$.  As seen by the solid curve in Figure~\ref{fig:poten_t00}, the symmetric shape of the potential eventually emerges after the quantum evolution of the $k$-integration.
It should be noted that this nontrivial feature has already been manifest in the original theory; see Eq.~\eqref{eq:Gamma_order2}.
The coefficient of $|Z|^2$ is positive leading to the minimum at $Z=0$.
In the deformed theory the ground state energy is given by $V_{k\to 0}(r=0)$ and
we draw the dot-dashed curve in Figure~\ref{fig:ELPA} to show the fRG results in the LPA in this way.
We see that the $\theta$-dependence of the ground state energy is quantitatively captured up to $\sim 2\pi$.

\begin{figure}
    \centering
    \includegraphics[width=0.7\textwidth]{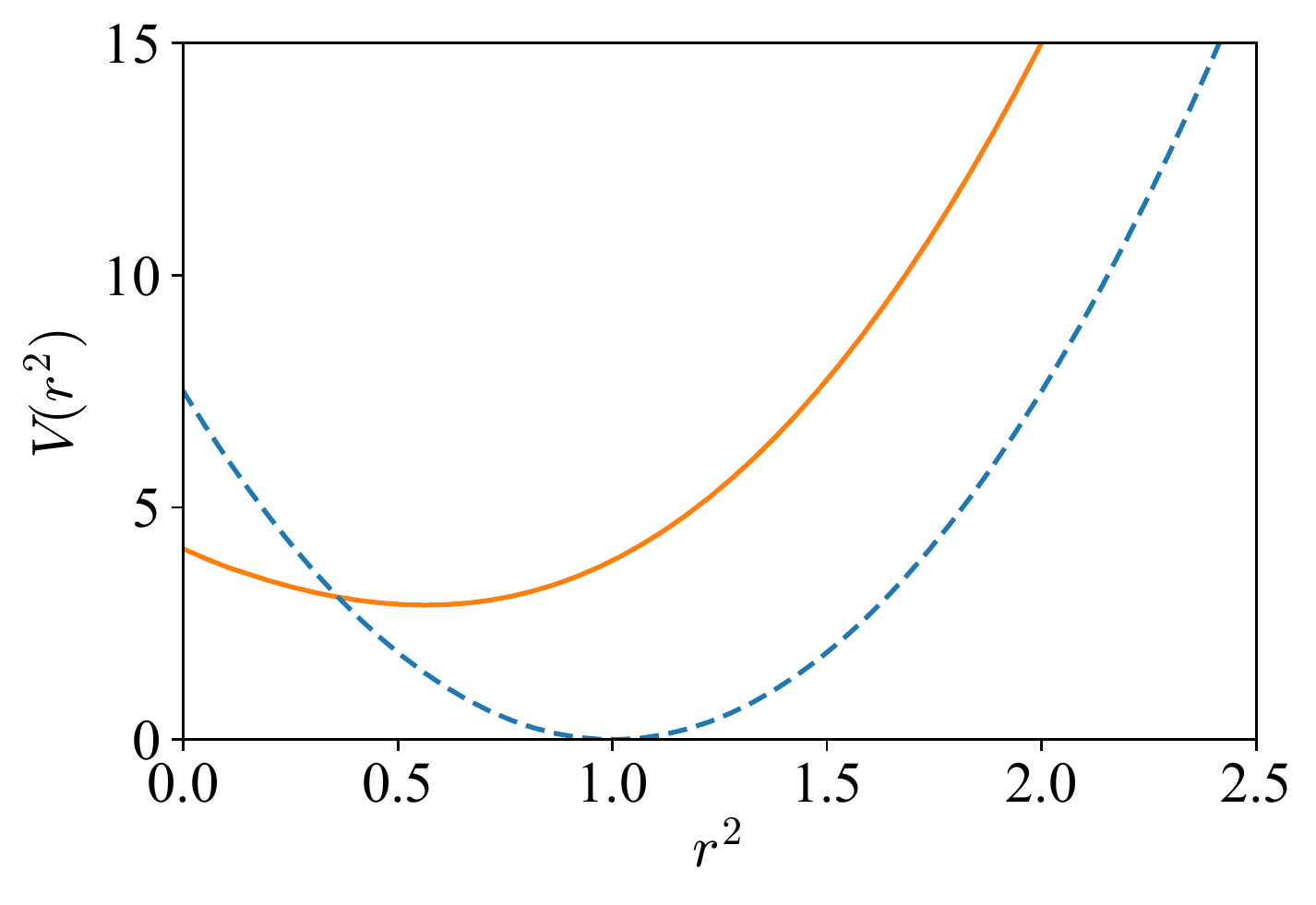}
    \caption{Effective potential at $\theta=3\pi$ evolved to $k\simeq 2.944$ where the evolution stops.  The dotted curve represents the initial effective potential as shown in Fig.~\ref{fig:poten_t00}.} \label{fig:poten_t3p}
\end{figure}

The dot-dashed curve in Figure~\ref{fig:ELPA} is terminated around $\theta\sim 2\pi$.  This is actually the second nontrivial feature we have encountered.
The fRG equation typically involves an energy denominator.  In the present case the denominator in Eq.~\eqref{eq:fRG_LPA} becomes vanishing for $mk^2 + 2V'=0$.
Usually such a singular point is avoided by the fRG equation itself due to the convexity of the potential.
We have carefully investigated the numerical calculation and have adjusted the step size of the $k$-integration in proportion to $mk^2 + 2V'$.
As long as $\theta$ is $\lesssim 2\pi$, not very close to the crossing point of the $n=0$ and the $n=1$ levels of the eigenenergies (see the solid and the dashed curves in Figure~\ref{fig:ELPA}), the numerical calculation can proceed to $k\to 0$ without difficulty.  For $\theta \gtrsim 2\pi$, however,
we have reached a conclusion that the $k$ evolution cannot avoid hitting the singularity and we should stop the integration.
Figure~\ref{fig:poten_t3p} shows an example of the potential at $\theta=3\pi$.
In this case we found that $k$ cannot go smaller than $\sim 2.944$.
The solid curve in Figure~\ref{fig:poten_t3p} does not exhibit convexity yet.
Thus, the fRG method in the LPA breaks down there, and the ground-state energy cannot be evaluated at all.
This feature is surprising, but understandable from Eq.~\eqref{eq:Gamma_order2} in the original theory again.
The coefficient of $|Z|^2$ approaches zero at the energy level crossing and it would go unphysically negative if Eq.~\eqref{eq:Gamma_order2} is forced to be applied for $|\theta|>\pi$ out of the validity range of the expression.

For completeness, we shall give numerical details here.
We discretized $r^2 \in [0, 2.5]$ with $50$ equally spaced points, i.e., $\Delta r^2 = 0.05$.
Then, we represent $V(r^2)$ on this grid and adopted the 5-point formulas to approximate $V'$ and $V''$ except for the edges.
At the edges we used the 3-point formulas.  For example, using a notation of $V[n] = V[n\Delta r^2]$, we calculated $V'$ from
\begin{equation}
    V'[n] = \frac{V[n-2] - 8V[n-1] + 8V[n+1] - V[n+2]}{12 \Delta r^2} + \mathcal{O}'((\Delta r^2)^4)\,,
\end{equation}
which is, at the edge $n=0$, replaced with
\begin{align}
    V'[0] &= \frac{-V[2] + 4V[1] - 3V[0]}{2\Delta r^2}
    + \mathcal{O}((\Delta r^2)^2)\,,\\
    V'[1] &= \frac{V[2] - V[0]}{2\Delta r^2} + \mathcal{O}((\Delta r^2)^2)\,.
\end{align}
We treated the upper edge at $r^2=2.5$ in the same way.
As we checked in the harmonic oscillator example,
the convergence to the correct answer as a function of increasing $\Lambda$ is logarithmically slow, and we numerically confirmed that $\Lambda=3000$ is large enough to reproduce the correct answer for $\theta=0$.
Then we performed the numerical integration with the 5th-order Runge-Kutta algorithm until $k = 1\times 10^{-3}$ with adaptive step size, $\Delta k$.

\subsection{Difficulty at \texorpdfstring{$\theta=\pi$}{theta=pi} and the 't~Hooft anomaly}
\label{sec:tHooft_deform}

One might think that the failure of the fRG approach in the LPA is a technical problem, but we would emphasize that this problem has a profound origin.
As summarized in Sec.~\ref{sec:spectrum_S1}, there exists the level crossing at $\theta=\pi$ (in the original theory, $g\to \infty$) and the most important property manifested there is the degeneracy of the ground state.
As we discussed in Sec.~\ref{sec:spectrum_S1}, this degeneracy at $\theta=\pi$ arises from the 't~Hooft anomaly in quantum field theory.
Usually, such degeneracy of ground states is circumvented due to the level repulsion, but it does not work in this case because two ground states have different $U(1)$ charges.

In our present attempt $g=30$ is still far from infinity, but the crossing of the $n=0$ and the $n=1$ levels near $\theta\sim 2\pi$ in Figure~\ref{fig:ELPA} is obviously traced back to the degeneracy at $\theta=\pi$ in the original theory.
In this case with $g<\infty$, the degeneracy is not associated with spontaneous symmetry breaking.
However, we note that the label $n$ characterizes the $U(1)$ charge of each state, and the level crossings of both cases occur between the states of different $U(1)$ charges, $n=0$ and $n=1$.
Let us point out that this is a prototypical example of the quantum phase transition between different symmetry-protected topological (SPT) states in quantum many-body physics.

It is a striking observation that, even though the fRG formalism itself could be nonperturbatively exact, the simple but common LPA could fail to describe such states related to topology.
Quantum phase transitions are so common in contemporary physics, and the LPA would be the first choice of approximation in the fRG calculation.
However, our analysis makes it clear that such a combination does not work properly for certain class of problems.
In the next section, we perform an analytic computation of the quantum effective action at $\theta=\pi$ to get better understanding on the failure of LPA.

Here, it would be fair to point out that the exact calculation of the effective action in Sec.~\ref{sec:effective_action_S1} already suggests the failure of the LPA when we take the $g\to \infty$ limit.
Both the kinetic term and the $\theta$ term in Eq.~\eqref{eq:Gamma_order2} has an extra factor $2$ compared with the classical action.
It suggests that a large wavefunction renormalization should be developed as $k\to 0$ under the fRG flow when $g$ is sufficiently large, but the LPA ansatz~\eqref{eq:Gamma_LPA} does not capture this feature.
In our above analysis, we take an intermediate value, $g=30$, so that the LPA still works well, while this value of $g$ is large enough to realize the level crossing.
Therefore, we do not think that this is the origin for the failure of the LPA in our computation.
Still, it is an interesting future study to improve the LPA to, e.g., the LPA$'$~\cite{Tetradis:1993ts,Blaizot:2005wd} for performing computations with larger values of $g$.

\section{Alternative formulation with the Villain lattice action}
\label{sec:Villain}

To understand better how the fRG in the LPA fails, it would be instructive to solve the problem in an alternative (and more rigorous) way.
We make use of the Villain lattice formulation for this purpose.
In this section we turn back to the original theory without the auxiliary parameter $g$, or at $g\to \infty$.

\subsection{Solving quantum mechanics on \texorpdfstring{$S^1$}{S1} with the Poisson summation formula}

We can solve the problem to find the same analytical answer on the discretized lattice.
We shall see that the lattice results coincide the ones in the continuum limit for $-\pi <\theta<\pi$, where the ground state is unique.
In addition, we will look more in details about the case with $\theta=\pi$.

We take the following form of the action with the lattice regularization:
\begin{equation}
    S = \sum_i \frac{m}{2a}
    (\Delta \phi_i - 2\pi A_i)^2
    - \im \frac{\theta}{2\pi}(\Delta \phi_i - 2\pi A_i)\,,
\end{equation}
which corresponds to the Euclidean Lagrangian in Eq.~\eqref{eq:L_original}.  Here, $i=1,\dots,N$ label the lattice sites and $a$ denotes the lattice spacing.
Thus, $\beta=Na$ is the period in the imaginary time direction.
The dynamical variables are $\phi_i\in \mathbb{R}$ corresponding to discretized $\phi$ and $\Delta \phi_i=\phi_{i+1}-\phi_i$ is introduced.
We note that with $A_i\in\mathbb{Z}$ this theory has gauge invariance under the following $\mathbb{Z}$-valued gauge transformation:
\begin{equation}
    \phi_i \mapsto \phi_i + 2\pi\lambda_i\,,\qquad
    A_i \mapsto A_i + \Delta\lambda_i\,.
\end{equation}
The gauge parameter is $\lambda_i\in\mathbb{Z}$, so that $\Delta\lambda_i = \lambda_{i+1}-\lambda_i\in\mathbb{Z}$.
Owing to this gauge symmetry we can choose $\lambda_i$ in the Villain gauge and restrict the dynamical variable range as
\begin{equation}
    -\pi \le \phi_i \le \pi\,.
\end{equation}
We note that the Villain gauge corresponds to the decomposition we made in Eq.~\eqref{eq:decomposition}.
In the same way as we saw from Eq.~\eqref{eq:bf_Poisson} to Eq.~\eqref{eq:af_Poisson}, we uses the Poisson summation formula and find the following expression up to an overall constant:
\begin{equation}
    \calZ = \sum_{\{n_i\}\in\mathbb{Z}^N} \int \Diff \phi\,
    \exp\biggl[ -\sum_i \frac{a}{2m} \biggl(
    n_i - \frac{\theta}{2\pi} \biggr)^2
    - \im \sum_i \Delta n_i\, \phi_i \biggr]\,.
\end{equation}
We can immediately perform the $\phi$-integration.  Then, it imposes a condition,
\begin{equation}
    \Delta n_i = 0\,.
\end{equation}
Therefore, we conclude $n:=n_1=n_2= \dots =n_N$.
Then the partition function reads:
\begin{equation}
    \calZ = \sum_{n=-\infty}^\infty
    \exp \biggl[ -\frac{\beta}{2m}\biggl(
    n - \frac{\theta}{2\pi} \biggr)^2 \biggr]\,,
\end{equation}
which recovers the ground-state energy correctly.
Interestingly, as advertised, the exact results are obtained even without taking the continuum limit.

\subsection{Constructing the quantum effective action near the level crossing}

Now let us consider the generating functional with source terms as defined by
\begin{equation}
    \calZ[J,J^\ast] = \int\Diff \phi\, \Diff A\;
    \exp\biggl( -S[\phi,A] + \rme^{\im \phi}\cdot J + \rme^{-\im \phi}\cdot J^* \biggr)\,.
\end{equation}
Here, $\rme^{\im \phi}\cdot J=\sum_i \rme^{\im \phi_i} J_i$.
In the same way as we did in the previous subsection,
we can reorganize the sum over $A$ using the Poisson formula to find,
\begin{equation}
    \calZ[J,J^\ast] = \sum_{\{n_i\}\in \mathbb{Z}^N}
    \int\Diff \phi\, \exp\Biggl[
    -\sum_i \frac{a}{2m}\biggl(n_i-\frac{\theta}{2\pi}\biggr)^2
    - \im \Delta n\cdot \phi +
    J\cdot \rme^{\im \phi} + J^\ast\cdot \rme^{-\im \phi} \Biggr]\,.
\end{equation}
We can perform the $\phi$-integration.  For simplicity let us expand the above expression in terms of $J_i$ and construct the quantum effective action perturbatively in terms of $J_i$ as we did for the continuum formulation in Sec.~\ref{sec:effective_action_S1}.

It is easy to see the analytical structure;  one $J_i$ and one $J_i^\ast$ fall down from the exponential and the $\phi$-integration leads to the Kronecker delta function as
\begin{equation}
    \calZ[J,J^\ast] = \calZ[0,0] + \sum_{j,k} J\cdot \tilde{G}\cdot J^\ast \,,
\end{equation}
where the matrix element of $\tilde{G}$ is given as
\begin{equation}
    \tilde{G}_{j,k} = \sum_n \rme^{-\beta E_n(\theta)}
    \biggl[ \Theta(j-k) \rme^{-(j-k)a (E_{n-1}(\theta)-E_n(\theta))}
    + \Theta(k-j) \rme^{-(k-j)a
    (E_{n+1}(\theta)-E_n(\theta))}\biggr]\,.
\end{equation}
This is a quite instructive form.
Previously, at $\beta\to\infty$, we assumed that the vacuum should be $|0\rangle$ for $-\pi < \theta < \pi$ and left only the $n=0$ contribution.
In the above expression it is already evident that the contribution from $n$ that minimizes $E_n(\theta)$ would dominate the sum over $n$ due to the overall exponential factor, $\rme^{-\beta E_n(\theta)}$.  Taking the continuum limit as well as the $\beta\to\infty$ limit, therefore, for $-\pi < \theta < \pi$ the above form simplifies as
\begin{equation}
    \tilde{G}(\tau) = \rme^{-\beta E_0(\theta)}\,
    G(\tau)\,,
\end{equation}
where $G(\tau)$ is given by Eq.~\eqref{eq:G(tau)}.
As $\calZ[0,0]$ is also dominated by $\rme^{-\beta E_0(\theta)}$, we get,
\begin{equation}
    \mathcal{W}[J,J^\ast] = \ln \mathcal{Z}[J,J^\ast]
    = -\beta E_0(\theta) + \ln ( 1 + J\cdot G\cdot J^\ast) \simeq -\beta E_0(\theta) + J\cdot G\cdot J^\ast\,.
\end{equation}
For the construction of $\Gamma[Z, Z^\ast]$, the rest of the procedures are just the same as we considered in Sec.~\ref{sec:effective_action_S1}.

Next, let us now focus on the level crossing of $n=0$ and $n=1$.
Important difference for $\theta=\pi$ is that two contributions with different $n$'s become comparable as $E_0=E_1$, and the effective action would change its form drastically.
We expand the Schwinger functional at the quadratic order in $J$ and $J^*$ as
\begin{align}
    \mathcal{W}[J,J^*]&=\ln \calZ[J,J^*]=\ln(\rme^{-\beta E_0}+\rme^{-\beta E_1})+J\cdot G_{\pi}\cdot J^*\nonumber\\
    &=\ln 2-\beta E_0(\theta=\pi)+J\cdot G_{\pi}\cdot J^*.
\end{align}
Here, we use $E_0=E_1$ at $\theta=\pi$. The first term denotes the ground-state degeneracy, the second one does the ground-state energy, and the last term does the connected $2$-point function.
The Green function $G_{\pi}(\tau)$ in the continuum limit is given by
\begin{align}
    G_{\pi}(\tau)&=\frac{\biggl(
    \Theta(\tau) \rme^{-\tau(E_{-1}-E_0)} + \Theta(-\tau) \rme^{\tau(E_1-E_0)} \biggr) + \biggl(
    \Theta(\tau) \rme^{-\tau(E_0 - E_1)}
    + \Theta(-\tau) \rme^{\tau(E_2-E_1)} \biggr)}{2}\nonumber\\
    &=\frac{\Theta(\tau) \rme^{-\tau/m} + \Theta(-\tau) \rme^{\tau/m}}{2}
    +1.
\end{align}
To obtain the last expression, we used $E_{-1}-E_{0}=E_{2}-E_{1}=1/m$ at $\theta=\pi$.
Now, we would like to solve $Z=G_{\pi}\cdot J^*$ but we here encounter the problem.
Instead of Eq.~\eqref{eq:diff_G}, $G_{\pi}(\tau)$ satisfies
\begin{equation}
    \left(-m\partial_\tau^2+{1\over m}\right)(G_{\pi}(\tau)-1)=\delta(\tau).
\end{equation}
As a result, we cannot solve $J^*$ in terms of $Z$ in the local way, and instead we have,
\begin{equation}
    \left(-m\partial_\tau^2+{1\over m}\right)Z(\tau)=J^*(\tau)+{1\over m}\int \diff \tau' J^*(\tau').
\end{equation}
Keeping $\beta$ to be large but finite, we get
\begin{equation}
    J^*(\tau)= \left(-m\partial_\tau^2+{1\over m}\right)Z(\tau)-{1\over m(\beta+m)}\int\diff \tau' Z(\tau').
\end{equation}
Formally performing the Legendre transformation with this result, we obtain that
\begin{equation}
    \Gamma[Z,Z^*]_{\theta=\pi}=-\ln 2+\beta E_0+\int \diff \tau\left(m |\dot{Z}|^2+{|Z|^2\over m}\right)-{1\over m(\beta+m)}\left|\int \diff \tau' Z(\tau')\right|^2,
\end{equation}
and we see that the quantum effective action is completely nonlocal.
We can further check that this nonlocality originates from the ground-state degeneracy.

Our result strongly suggests that any kind of ansatz for $\Gamma_k$ in the local form is inappropriate to describe the level crossing phenomena of the ground states.
This poses a serious question on the practical applicability of the fRG to this system beyond the level crossing point, and we need to think of a nonlocal ansatz of the effective action in the future study to tackle this problem with fRG.


\section{Conclusions}

We have investigated the $\theta$-vacuum structure of the quantum mechanical system on $S^1$ with special emphasis on the applicability of the fRG formulation.
Since the fRG equation is expressed in the functional differential equation, it seems that any topological term which is unchanged under continuous differentiation would be dropped off from the fRG formulation.
The quantum mechanical system offers an appropriate test ground to reveal possible machinery of how the topological $\theta$ term could affect physics within the framework of the Wetterich equation.
Its simplicity allows us to scrutinize the energy spectrum in the canonical quantization and the path-integral approach.
In particular, the $\theta$-dependence of the eigenenergies and the associated degeneracy of the ground state at $\theta=\pi$ are demonstrated transparently.
This system exhibits a prototype of the mixed 't~Hooft anomaly between the $U(1)$ symmetry and the charge conjugation symmetry, which explains doubly degenerate ground states at $\theta=\pi$.
In addition to the anomaly content,
this simple system at finite $\theta$ suffers from the sign problem and poses an interesting question on the existence of the quantum effective action when the sign problem ruins the convexity.
The generating functional, $\mathcal{W}[J, J^*]$, takes a complex value for $\theta\not\in 2\pi\mathbb{Z}$ in our quantum mechanical system.
Therefore, we would like to point out a potential pitfall that the fRG may not be completely free from the notorious sign problem.
In our case, thanks to the unbroken continuous symmetry, we can construct the effective action as a formal power series.  In general problems in quantum field theories, this issue is more subtle and deserves careful investigations in the future.

The $2\pi$-periodicity in the target space $S^1$ would be incompatible with the fRG equation as it is.
The problem is circumvented by embedding the $S^1$ target space into $\mathbb{R}^2$ with the wine-bottle potential $\propto g$.
With $g<\infty$, on the one hand, configurations can distribute over $\mathbb{R}^2$ instead of $S^1$, and the winding number is no longer topological.
On the other hand, the original theory on $S^1$ is recovered in the $g\to\infty$ limit.
This deformed theory with finite $g$ is still a simple quantum mechanical system so that the energy spectrum is obtained without difficulty.
We examined the $\theta$-dependent ground state energy using the Wetterich equation.
We truncated the effective average action in the LPA, i.e., at the leading order of the derivative expansion.
The Wetterich equation in the LPA gives the correct results until the energy level crossing occurs as shown in Figure~\ref{fig:ELPA}.
When the level crossing happens, we found that
the LPA fails to work at all and the RG flow is stopped.  Although the fRG formalism itself is nonperturbatively exact, it is interesting that such a simple model challenges the applicability of the LPA calculation.
Moreover, the alternative formulation based on the Villain lattice action raises more serious problems.
Using the Villain formulation, we can write down the effective action at $\theta=\pi$, where the level crossing of ground states occurs.
Our observation is that, when the energy level crossing occurs, the effective action takes the nonlocal form, and thus it suggests that any kind of derivative expansion does not work in this problem.

In this work, we aim to point out a problem in the clearest way that the fRG approach could encounter difficulty in featuring the $\theta$-vacuum structure correctly.
It is intriguing to go beyond the LPA and,
as the Villain lattice analysis suggests, to try to describe the level crossing behavior with the fully nonlocal effective action.
In summary, the $S^1$ quantum mechanics is a simple but very useful model to delve into topological contents that should be retained in the fRG formulation.

\acknowledgments

The authors thank
Mithat \"{U}nsal
who asked a question on the topological term in the fRG framework at a workshop in Oberw\"{o}lz.
KF was inspired by his critical question.
This work was partially supported by Japan Society for the Promotion of Science
(JSPS) KAKENHI Grant
Nos.\ 18H01211 (KF), 21J11298 (TS), and JSPS KAKENHI Grant-in-Aid for Research Activity
Start-up, 20K22350 (YT).

\bibliographystyle{JHEP}
\bibliography{./QFT.bib,./refs.bib}

\end{document}